\documentclass[a4paper,11pt]{article}
\pdfoutput=1 

\usepackage{jheppub} 

\usepackage[T1]{fontenc} 
\usepackage{braket}
\usepackage{multirow}
\usepackage{amsmath}
\usepackage{physics}
\usepackage[table]{xcolor}
\usepackage{adjustbox}
\usepackage{sidecap}

\newcommand{\lb}{\left(}
\newcommand{\rb}{\right)}

\title{\boldmath Hartree-Fock all-heavy $c$, $b$ multiquarks 
\\
and constraints on new top-sector physics}

\author{Alejandro Alonso-Valero,}
\author{Daniel Berzal-Rozal\'en,}
\author[1]{Felipe J. Llanes-Estrada\note{On leave at the Theory Department of CERN, 1211 Geneva, Switzerland. Corresponding author.},} 
\author{\\Mario Camilo Pardo,}
\author{and Clara Peset}
\affiliation{Depto. F\'{\i}sica Te\'orica \& IPARCOS, Univ. Complutense de Madrid, Plaza de las Ciencias 1,
28040 Madrid, Spain}

\emailAdd{alealo10@ucm.es}
\emailAdd{daberzal@ucm.es}
\emailAdd{fllanes@fis.ucm.es}
\emailAdd{mariocap@ucm.es}
\emailAdd{cpeset@ucm.es}

\abstract{We deploy the Hartree-Fock approximation for all-heavy quark hadrons, including quarkonium, baryons, tetraquarks, pentaquarks, dibaryons and up to the 12-body dibaryon-antidibaryon which completely fill the $1s$ orbital, in a unified manner, with the spinless LO Coulomb interaction and beyond. After treating the $c$ and $b$ quarks in various combinations, we delve a bit longer on 
$t$-quark bound states. We extend the negative result of Kuchiev, Flambaum and Shuryak on the 12-body topball 
to now include the NLO QCD potential. We find that none of the examined multitop states should have binding energy exceeding their width in the Standard Model.  Additional new-physics interactions  can then be constrained by experimental searches for these bound states, assuming they are not detected, at a level somewhat less stringent than standing HEFT bounds, but with different systematics.}

\preprint{IPARCOS-UCM-2024-048}

\begin{document}

\maketitle
\flushbottom

\section{Introduction}\label{intro}

A lasting theme of hadron spectroscopy is the existence and properties~\cite{Hanhart:2019isz} of resonances which evade classification as conventional $q\bar{q}$ mesons or $qqq$ baryons. It is evident that atomic nuclei do exist, so there is no real surprise in finding that other quark flavours may also form molecule-like entities bound by long-distance pion exchange, but exploring which ones do exist and what their properties are has been a rewarding enterprise~\footnote{
The first example was Belle's X(3872)~\cite{Belle:2003nnu}, believed to include a $uc\bar{u}\bar{c}$ state in its wavefunction.}. 
The more interesting question for many spectroscopists is whether ``compact'' multiquark~\cite{Jaffe:1977cv} (i.e. tetraquark, pentaquark...) structures, or even glueballs~\cite{Llanes-Estrada:2021evz}, do exist while being bound by short-range colour forces. This has been notoriously difficult for several reasons. One is the possible rearrangement of the whole structure into color-singlet subclusters which again gives them the appearance of meson-meson, meson-baryon, baryon-antibaryon, etc. molecules~\cite{Molina:2009ct}. How can the short- and long-distance effects be disentangled?

All-heavy-quark systems are different in that the distance between heavy quarks scales as the inverse mass $1/m$ and so they are more sensitive to short-range forces, particularly in their ground states (potential Non Relativistic QCD formalizes this insight~\cite{Brambilla:1999xf}). They therefore offer access to ``contact'' interactions, a kind of ``colour nuclear physics''. On the minus side, their production is challenging given the cross-section suppression respect to conventional hadrons.

In the last years, all-heavy hadrons are starting to look more likely (as baryons and probably molecular systems with two heavy quarks, $qQQ$ and $qqQQ$ have been reported~\cite{Brambilla:2019esw} and are under intense theoretical scrutiny~\cite{Anwar:2017toa,Liu:2019zoy,Berwein:2024ztx}), so there is immense theoretical interest in their study. Reasonable questions that we would like to see answered are whether any multiquark states are below their fall-apart threshold into singlet baryons and mesons; why only a few key quantum numbers seem to appear, while a large part of the quantum numbers which one can form with several particles only form continua; and how their properties deform with quark mass to match the lighter combinations which form non-heavy hadrons, so we learn about the nuclear forces in all regimes.

While all-heavy baryons are, mostly, yet to be found (saliently, the $\Omega_{ccc}$ triply charmed baryon~\cite{Llanes-Estrada:2011gwu} has not yet been reported), there is a structure, $X(6900)$, near the $J/\psi J/\psi$ threshold, claimed by LHCb  and confirmed by the ATLAS and CMS collaborations. Five further new possible structures populate the $6.2-7.2$ GeV range \cite{LHCb:2020bwg,CMS:2023owd,ATLAS:2023bft}. Any of these could be a candidate for an all-heavy tetraquark, but note that many hadron phenomena, such as the opening of thresholds, triangle singularities and so on can also produce peaking structures in a spectrum~\cite{Bugg:2011jr,Wang:2017mrt,Abreu:2020jsl}.
None with a higher number of quarks has been detected up to date, but developing theoretical methods which can handle an arbitrarily large number of heavy quarks does not seem far-fetched now. 

Theoretical research on fully-heavy tetraquarks has been conducted using a wide variety of methods such as QCD sum rules~\cite{Tang:2024zvf}, Green's function Monte Carlo methods~\cite{Assi:2023dlu},
 Bethe-Salpeter equations~\cite{Heupel:2012ua,Liu:2024pio} and various quark models~\cite{Galkin:2023wox}.
Well known is a lattice calculation~\cite{Hughes:2017xie}
which failed to find evidence of all-beauty tetraquarks.

We contribute to this research line and in the first part of this work (sections \ref{preliminaries} to \ref{section:results2}) we adapt the Hartree-Fock (HF) method from atomic physics, ideally suited for a system in which many bodies interact via a nonrelativistic potential with pair-creation suppressed by $1/m$ factors.
We will extensively use the central color Coulomb interaction, but also the NLO potential natural to the pNRQCD framework, and only briefly a spin-spin interaction.

 We will address the simplest heavy-quark systems where all particles occupy the ground-state $s$-wave orbital. With 3 colors and spin $1/2$, this allows for up to 6 quarks and 6 antiquarks of each flavour. 

That ``dodecaquark'' state, bound perhaps by Higgs exchange instead of by Chromodynamics forces, was proposed in the $t$-quark sector by Froggatt and Nielsen~\cite{Froggatt:2003tu,Froggatt:2004bh} in thinking about the Standard Model (SM) parameters,
before the discovery of the Higgs boson with $m_h\approx$ 125 GeV. They dubbed the state the T-ball and estimated it to be very bound, even reaching null mass with modifications of the top parameters and a light Higgs. 

Kuchiev, Flambaum and Shuryak~\cite{Kuchiev:2008fd} carried out a more precise computation using, precisely, a Hartree-Fock method. They found that, in order for the T-ball to be bound by the Yukawa interaction alone, the Higgs mass could not exceed about 27 GeV, concluding that this interaction does not produce new bound states. Analytical bounds implemented by Richard~\cite{Richard:2008uq}
concur in that there is a maximum scalar-boson mass between around 30 and 40 GeV beyond which the potential supports no bound state.

The top quark~\cite{Dawson:2003uc}, being the heaviest known elementary particle with a mass $\sim$ 170-175 GeV, decays weakly into $b + W^{+}$ with a width $\Gamma_t\sim$ 1.3 GeV \cite{golowich}. Due to this large decay rate, there is no time for the top quark to form bound-state hadrons: its life time $\tau = \Gamma_t^{-1} \sim 10^{-25}$s is 10 times smaller than the hadronization time scale $\Lambda_{QCD}^{-1} \sim 10^{-24}$s, although the QCD interaction is binding. 

A question that we can answer in the second part of this work (sections \ref{sec:tpotentials} to \ref{sec:numerictop}), upon deploying the Hartree approximation, is whether larger binding energies and thus faster hadronization times can be achieved by placing an increasing number of top-quarks into the same ground state and let them interact by an attractive force. 

We will examine this problem  within the Standard Model but also in the presence of new-physics interactions. This addresses whether the T-ball could still be very bound 
as Froggatt and Nielsen would have it, given the constraints on new physics that have already been put forward (we will find it is not), 
and, reversing the logic, what constraints on the new-physics parameters can be expected from searches for toponium and multitop states such as the T-ball.
It is important to address the $t$-quark sector from all points of view because, it being the heaviest produced particle, it is quite strongly coupled to the electroweak symmetry breaking sector of the Standard Model~\cite{Quezada-Calonge:2022lop,Quezada-Calonge:2022rsc} and one of the most sensitive probes of eventual high-scale new physics.

The new-physics interactions will be parametrized with four-quark operators in the notation of the Higgs Effective Field Theory (HEFT) Lagrangian~\cite{Brivio:2016fzo}.  

Six states will be considered: toponium ($t\bar{t}$), the top-baryon ($ttt$), the tetratop ($tt\bar{t}\bar{t}$), pentatop ($tt\bar{t}t\bar{t}$), hexatop ($ttt-ttt$) and the T- ball ($6t + 6\bar{t}$). Several interactions will be considered: gluon exchange ($V\sim 1/r$), Higgs and Z exchange ($V\sim e^{-m_{H}r}/r$) and the four new-physics contact interactions from Table \ref{tab:operators_sb} ($V \sim \delta(r)$).

\section{Reprise of minimum Hartree-Fock theory}\label{preliminaries}
While Hartree-Fock theory, not explicitly Lorentz invariant, is not so often treated in high-energy physics (with notable exceptions, e.g.~\cite{Witten:1979kh}) it  remains a basic tool of bound state physics and can lead to rather direct computations in spectroscopy in the nonrelativistic limit, taking into account the difficulty of treating many-body systems in a totally invariant manner~\cite{Kvinikhidze:2014yqa}.

\subsection{Basic equations}

Starting by the Hartree approximation to the many-body wavefunction, the trial wavefunction is a Hartree product of spin orbital wave functions for individual particles
\begin{equation}
	\Psi (q_1,...,q_N) = \psi_1(q_1)... \psi_N(q_N)
	\label{eq:prod_spinorbitals}
\end{equation}
(where $q = (\mathbf{x},s)$ contains the spatial and spin coordinates of an electron, a nucleon in a nucleus or a quark, and in that case, also flavour and colour). This solution wave function may also be referred as a \emph{Hartree product}. It has an uncorrelated probability distribution, since the probability density of finding the i-th particle in the volume element $d\mathbf{x}_i$ centered at $\mathbf{x}_i$ can be written as a product of the one-particle probability densities,
$
	\rho (q_1,...,q_N) = \left|\psi_1(q_1)\right|^2... \left|\psi_N(q_N)\right|^2
$.
Indistinguishability suggests to improve Eq.~(\ref{eq:prod_spinorbitals}) to an antisymmetric fermion wavefunction, which is achieved by the Slater determinant
\begin{equation}
	\Psi (q_1,...,q_N)= \frac{1}{\sqrt{N!}}
	\mathrm{Det}(\psi_i(q_j)) \label{Slaterdet}
\end{equation}
introducing correlations such as
\begin{equation}
	\rho(q_1, q_2) = \frac{1}{N(N-1)}	\sum_{k,l}\left[\left|\psi_k(q_1)\right|^2\left|	\psi_l(q_2)\right|^2-\psi_k^*(q_1) 	\psi_k(q_2) \psi_l(q_1) \psi_l^*(q_2)\right] \ .
	\label{eq:correlated_prob}
\end{equation}
Following the variational-approach philosophy, the energy will be obtained by optimization under $\psi \rightarrow \psi + \delta\psi$:
\begin{equation}\label{variational}
    \delta \left[ \bra{\Psi}H\ket{\Psi} - \sum_i \epsilon_i \bra{\Psi}\ket{\Psi} \right] = 0\ .
\end{equation}
To proceed we need to specify the Hamiltonian.  At short distance the strong interaction can be approximated at Leading Order (LO) single-gluon exchange, leading to the Coulomb potential except for a color factor. In natural units, for two quanta at distance $r$,
\begin{equation}
    V(r) = \pm f\frac{\alpha_s}{r}\equiv \kappa_c\frac{\alpha_s}{r},
    \label{eq:qcd_LOpotential}
\end{equation}
where $\alpha_s$ is the strong interaction coupling constant and $f$ is the color factor. It is positive for $qq$ and $\Bar{q}\Bar{q}$ interactions and negative for $q\Bar{q}$. A difference with atomic/molecular physics is the absence (for all-heavy quarks) of the nuclear terms in Born-Oppenheimer approximation, third of the analogous Hamiltonian
\begin{equation}
    H_{BO} =\sum_{i=1}^N\frac{p_i^2}{2m}+
    \frac{\alpha}{2} \sum_{i,j=1;i\neq j}^N \frac{1}{\left|\mathbf{r}_i -\mathbf{r}_j\right|} - \alpha
    \sum_{n=1}^K\sum_{i=1}^N \frac{Z_n}{\left|\mathbf{r}_i -\mathbf{R}_n\right|}\ .
    \label{eq:BO_hamiltonian}
\end{equation}

Also, to calculate the color factor in Eq.~(\ref{eq:qcd_LOpotential}) we add a color part to the wavefunction of the system constituents. As such,
$    \psi=\phi(\mathrm{spatial})\otimes\chi(\mathrm{spin})\otimes\sigma(\mathrm{flavour})\otimes\xi(\mathrm{color})
$
and the Hartree-Fock equations now read $\mathcal{F}\psi_k = \epsilon\psi_k$ with the Fock operator
\begin{multline}
	\mathcal{F}\psi_k(q) = -\frac{1}{2m}\nabla^2 \psi_k(q) + \kappa_c\sum_{l=1}^N\int{dq'\left|\phi_l(q')\right|^2 \frac{\alpha_s}{\left|\mathbf{r}-\mathbf{r'}\right|}\psi_k(q)} -\\ -\kappa_c\sum_{l =1}^N\delta(\chi_l,\chi_k)\delta(\sigma_l,\sigma_k)\int{dq'\phi_l^*(q')\frac{\alpha_s}{\left|\mathbf{r}-\mathbf{r'}\right|}\phi_k(q')\psi_l(q)}.
	\label{eq:Hartree_Fock_QCD}
\end{multline}

It is customary, and useful, to simplify the three-dimensional Laplacian, to employ the reduced wavefunction $P(r)\equiv rR(r)$ instead of $R$ itself. If the potential is also expanded in spherical harmonics, the angular part of the Hartree-Fock equation
can be computed beforehand, and only the radial integrals are dealt with during iteration. We do this numerically,
in general, but for the Coulomb electric potential the expansion can be  analytically given,
\begin{equation*}
    \frac{1}{|\mathbf{r}-\mathbf{r}'|}=\sum_{\ell=0}^\infty\sum_{m_\ell=-\ell}^\ell\frac{4\pi}{2\ell+1}\frac{r_<^\ell}{r_>^{\ell+1}}Y_{\ell m_\ell}^*(\Omega)Y_{\ell m_\ell}(\Omega'),
\end{equation*}
with  $r_>=\mathrm{max}(r,r')$ and $r_<=\mathrm{min}(r,r')$. As we will only address quark systems up to $N=12$, only the 
$l=0=m_l$ term will be necessary, bringing about great simplification. 
For an arbitrary potential 
\begin{equation}
    V(|\mathbf{r}-\mathbf{r}'|)=\displaystyle\sum_{\ell=0}^\infty\sum_{m_\ell=-\ell}^\ell C_{\ell m_\ell}(r,r')Y_{\ell m_\ell}^*(\Omega)Y_{\ell m_\ell}(\Omega')\ ,
\end{equation}
so that for purely spherical orbitals in the fundamental state only one kernel is needed,
\begin{equation}
    C_{00}(r,r')=\frac{1}{4\pi}\int V(|\mathbf{r}-\mathbf{r}'|)d\Omega d\Omega'\ .
    \label{c00}
\end{equation}

The typical Hartree integral,
\begin{equation}
    \mathcal{H}(r)=\alpha\int_0^\infty P^2(r')\frac{1}{r_>}dr'\ ,
    \label{Vit}
\end{equation}
is then discretized and computed by the trapezoidal rule, for example. Discretizing the Laplacian's second radial derivative,
the molecular HF equation becomes (as there is only one orbital, $P$, the exchange Fock term takes a form very similar to the Hartree one)
\begin{equation}
    -\frac{1}{2}\frac{P_{n+1}-2P_n+P_{n-1}}{h^2}-\frac{Z\alpha}{r_n}P_n+\sum_{j\neq i}\mathcal{H}_nP_n-\sum_{j\neq i}\delta_{m_{s_i}m_{s_j}}\mathcal{H}_nP_n=\varepsilon_i P_n\ .
\end{equation}
Some modifications are necessary when addressing flavor and colour in the $SU(3)$ gauge theory.
For example, incorporating flavour $f=c,b$ additionally to the spin states
$\chi=\chi_+,\chi_-$, the eigenvalue problem is
\begin{equation}
    -\frac{1}{2m_i}\frac{d^2P_i}{dr^2}\chi_i+\sum_{j(\neq i)}\mathcal{V}_{jj}(r)P_i\chi_i-\sum_{j(\neq i)}(\chi_i^\dagger\chi_j)(f_i^\dagger f_j)\mathcal{V}_{ij}(r)P_j\chi_j=\varepsilon_i P_i\chi_i\ ,
    \label{altpot}
\end{equation}
where
\begin{equation*}
    \mathcal{V}_{ij}(r)=\frac{1}{4\pi}\int C_{00}(r,r')P_i(r')P_j(r')dr'\ .
\end{equation*}
Only if two quarks share the same flavour $m_i=m_j$ do we assign them to the same orbital $P_i=P_j$.
For two different flavours, as an additional variational improvement, we solve two coupled Hartree-Fock equations, one for bottom (typically a more compact orbital) and one for charm (a more relaxed, extended one).

Note that the color state of the entire system must be a confined singlet under  $SU(3)$. That is, the system will be in the physical spectrum only if the color state product of its constituents decomposes to an irreducible representation with dimension equal to one~\cite{Lundhammar:2020xvw}. To see the combinations that result in a color singlet it is convenient to decompose the products SU(3) fundamental/antifundamental representations in irreducible ones, 
\begin{equation}
\ket{q\Bar{q}}:3\otimes\Bar{3}=1\oplus8\nonumber\ \ \ \ \ 
\ket{qq}:3\otimes3=\Bar{3}\oplus6\label{eq:twoquarks_su3}\ \ \ \ \  \ket{\Bar{q}\Bar{q}}:\Bar{3}\otimes\Bar{3}=3\oplus\Bar{6}\nonumber
\end{equation}
which we will assiduously use to form more complex states.

Once the HF equations are solved, Koopman's theorem yields the total energy~\cite{fisatom}
\begin{equation}
    E=\sum_i\varepsilon_i-(J-K)
    \label{energy}
\end{equation}
with $J$ and $K$, respectively, the Coulomb integral and, if appropriate, the identical-particle exchange one,
\begin{equation*}
    J=\alpha\sum_{i<j}\int\frac{|\Phi_i(\mathbf{r},\sigma)|^2|\Phi_j(\mathbf{r}',\sigma')|^2}{|\mathbf{r}-\mathbf{r}'|}d^3\mathbf{r}d^3\mathbf{r}'=\alpha\sum_{i<j}\int P^2(r)P^2(r')\frac{1}{r_>^{}}drdr'\ ,
\end{equation*}
\begin{equation*}
    K=\alpha\sum_{i<j}\int\frac{\Phi_i^*(\mathbf{r},\sigma)\Phi_j(\mathbf{r},\sigma)\Phi_j^*(\mathbf{r}',\sigma')\Phi_i(\mathbf{r}',\sigma')}{|\mathbf{r}-\mathbf{r}'|}d^3\mathbf{r}d^3\mathbf{r}'=\alpha\sum_{i<j}\delta_{m_{s_i}m_{s_j}}\int P^2(r)P^2(r')\frac{1}{r_>^{}}drdr'\ .
\end{equation*}
Eq.~(\ref{energy}) can be evaluated once some version of Eq.~(\ref{altpot}) has been solved  by numerical iteration, to which now we turn.
\subsection{Hartree-Fock computer program}
In this subsection we recall the general structure of a Hartree-Fock computer program.
\begin{enumerate}
    \item \textbf{Input functions:} Adapt the physical problem to the variational method, deciding the number of orbitals $N$ and the discretized $V_k, V_{kl}$ potentials. The starting functions for the iteration are typically hydrogenoid.
    \item \textbf{Calculate grid and introduce an initial trial ground state:} we have used two grids, one with $n$ Gaussian partitions of 20 points each which complicates the calculation of the Laplacian, while delivering more efficient integration, and an  equispaced grid with a simple Laplacian but less efficient integration.
    \item \textbf{Calculate matrices:}
    The uncoupled one-body Hamiltonian matrix does not depend on the ansatz wavefunction, so it is computed before entering the SCF loop.  To obtain the Laplacian's matrix, Fornberg's algorithm for calculating the weights for any order of derivative on an arbitrary grid in one dimension is considered \cite{fornberg1,fornberg2}. To avoid numerical errors at the grid's boundaries, no more than 6 weights per point should be considered. By calculating centered derivatives for the bulk of the grid, and advanced and retarded for the boundaries, the Laplacian is represented by a banded matrix. In an equispaced grid it reduces to the standard finite-difference radial second derivative.
\item \textbf{Loop to reach self consistency:} Once all kernels are built, the self consistent field method (SCF) is applied by finding new orbitals until convergence is met.
    \begin{enumerate}
        \item \textbf{Calculate the two-body interaction matrix:} The Coulomb contribution is stored in a diagonal matrix $J_{ii} = \sum_{l=1}^N\int{dq'\left|\psi_l(q')\right|^2 V_{kl}(\mathbf{r_i},\mathbf{r'})}$ that will be multiplied by $\psi_i(\mathbf{r}_i)$. Exchange contributions correspond to non-diagonal matrices.  
   \item \textbf{Orbital update step:} Diagonalisation of the summed Fock matrix yields a lowest eigenvalue $\varepsilon$ whose eigenvector represents the new orbital to be considered in the next SCF step.
    \item \textbf{Check for convergence:} a slowdown parameter is often used to limit the orbital advance between iterations. \cite{Thijssen}.
    \end{enumerate}
    \item \textbf{Calculate the ground state energy:} with the help of Koopman's theorem, Eq.~(\ref{energy}).
    
\end{enumerate}
\subsection{Checks: Helium and Positronium}
Before embarking on the computation of heavy-quark systems, we run quick checks for the two-electron Helium atom and the $e^-e^+$ ground states, employing just one spatial (1s) orbital as we will apply to heavy quarks.

Starting by Helium,  we separate the spatial and spin components $\psi_l(q)=\phi(\mathbf{r})\chi_l(s)$,  with $\phi(\mathbf{r})=R(r)Y_{00}=\frac{1}{\sqrt{4\pi}}R(r)$. The sum over spin contributes two terms to the Hartree kernel.
An analog treatment of the Fock term shows that it vanishes for $l\neq k$ and coincides with Hartree's for $l=k$, so that it cancels one of the two terms, leaving the other, and no factors. Performing now the angular integration, the HF equation is
\cite{franciscoblanco},
\begin{equation}
    \mathcal{F}R_k(r)=-\frac{1}{2m_er}\frac{d^2}{d r^2}(rR_k(r))-\frac{2\alpha}{r}R_k(r)+\int_0^\infty d r' r'^2R^2(r')\frac{\alpha}{\mathrm{max}(r,r')}R_k(r)
\end{equation}
which can be iteratively solved for $R_k$ or, preferably, for $P(r)\equiv rR(r)$, in order to avoid spurious numerical divergences at $r\rightarrow0$.

Once the iteration converged, the energy is given  (as there are $2\,e^-$s), from Eq.~(\ref{energy}), by $E=2\varepsilon-J$ or, in terms of the one-body terms $h$, equivalently by $E=\varepsilon_k+\mel{\psi_k}{h}{\psi_k}$.

We have run two independent programs (differing in the treatment of the radial grid) which yield similar energies, as in table~\ref{EHe} in great agreement (for such a simple variational method) with the experimental value (-79.0 eV with negligible uncertainty)\cite{heliumgs} and with other computational methods \cite{Thijssen}.

\begin{table}[h]
\centering
\caption{Ground state energy (total electron binding energy) of the neutral Helium atom, in eV.}
\begin{tabular}{|c|c|c|c|}
\hline
This work, HF1 & This work, HF2 & Thijssen~\cite{Thijssen}, HF & Experimental value \cite{heliumgs}\\ \hline
-77.9 & -78.57 & -77.69  & -79.01 \\ \hline
\end{tabular}
\label{EHe}
\end{table}

In fact our computation seems closer to data than the textbook's. This is likely a variational feature: whereas that earlier computation uses a basis with a few wavefunctions and relaxes the coefficients, we employ a discretization of the $r$ variable with hundreds to thousands of points. 
This can be thought of as a sequence of variational parameters which are the values of the $P$ orbital at each $n^{\rm th}$ value of $r$, providing a much more ample variational basis.

We then turn to positronium, which is a good testing ground for quarkonium-type calculations as it lacks a nucleus anchoring the atom. Because $e^-$ and $e^+$ are distinguishable, the Fock operator only includes the independent-particle Hamiltonian and the two-body Coulomb Hartree term, 
\begin{equation}
	\mathcal{F}\psi_k = -\frac{1}{4m_e}\nabla^2\psi_k(q) - \int{d^3\mathbf{r'}\left|\phi(\mathbf{r}')\right|^2\frac{\alpha}{\left|\mathbf{r}-\mathbf{r'}\right|}\psi_k(q)}
 \label{eq:fock_positronium}
\end{equation}

The lack of that anchoring heaviest mass requires setting the origin of coordinates at some fixed point. 
As a consequence, spurious momentum is included from not separating centre-of-mass motions. This can be remedied only partially at the one-particle level, by heuristically correcting each kinetic energy by a factor $-p^2_i/(2mN)$, where $N$ is the number of particles \cite{caltech_cmcorrection}. 
We are compelled to include an uncertainty which is $50\%$ of the correction (a more sophisticated treatment would approximately subtract the center-of-mass recoil~\cite{Hodges:2024awq,DeGregorio:2021dsr} given by an $N$-body operator). 
As implemented, both the correction and its uncertainty diminish as the number of bodies in the system is increased, since recoil is suppressed by $1/\sum_i m_i\propto \frac{1}{N}$.

Again factorizing the wavefunction, Fock's equation for the radial orbitals reads
\begin{equation}
    \mathcal{F}R_k(r)=-\frac{1}{4m_er}\frac{d^2}{d r^2}(rR_k(r))-\int_0^\infty d r' r'^2R^2(r')\frac{\alpha}{\mathrm{max}(r,r')}R_k(r).
\end{equation}
The exact value of positronium's ground state is widely known, and it can be easily deduced at leading order (LO) in the electric charge from $-\frac{1}{2}\alpha^2\mu=-\frac{1}{4}\alpha^2m_e\approx-6.8$ eV. The program yields a binding energy of $(-5.9\pm 1.6)$ eV. This result suggests that the uncertainty assigned to the kinetic energy correction might be overestimated, but we will keep it at this conservative value for the rest of the article.

\subsection{Two-body problem:\\  Schr\"odinger equation vs. Hartree-Fock variational approximation}
Since we will run the Hartree-Fock method for all multi-top quark systems later, it is worth delving a few lines on a comparison with the one case where an exact solution can be compared with, to understand the difference.
This is the two body problem, which can be exactly solved by the separation of the center of mass and relative particle coordinates,
\begin{eqnarray}
\nabla^2_{CM}=\nabla^2_1+\nabla^2_2+2\mathbf{\nabla}_1\cdot\mathbf{\nabla}_2\\
\nabla^2=\frac{1}{4}\left(\nabla^2_1+\nabla^2_2-2\mathbf{\nabla}_1\cdot\mathbf{\nabla}_2\right)\ .
\end{eqnarray}
which yields, of course,
\begin{equation}
    -\frac{1}{2M}\nabla^2_{CM}\psi-\frac{1}{2\mu}\nabla^2\psi-\frac{\alpha}{r}\psi=E\psi\ .
\end{equation}
The ground state solution is the usual exponential
\begin{equation}
    \phi(\vec{r})=\frac{2}{a^{3/2}}e^{-r/a}Y_{00}(\Omega)=\frac{1}{\sqrt{\pi a^3}}e^{-r/a}\ ,
    \label{ansatze+e-}
\end{equation}
with $a=2a_0=2/(\alpha m_e)$ twice the positronium Bohr radius, and where the spins  couple to a singlet. 

On the other hand, if we substitute as HF ansatz Eq.~(\ref{ansatze+e-}),
the Hartree and exchange integrals become
\begin{align*}
    \mathcal{V}(r)&=\alpha\int d^3\mathbf{r}' \frac{\left|\phi(\mathbf{r}')\right|^2}{\left|\mathbf{r}-\mathbf{r}'\right|} = \frac{4}{a^3}\left[\frac{1}{r}\int_0^r dr' r'^2e^{-2r'/a} + \int_r^\infty dr' r'e^{-2r'/a}\right]=\frac{1}{r} \left(1-e^{-\frac{2 r}{a}} \right)-\frac{1}{a}e^{-\frac{2 r}{a}}
\end{align*}
\begin{align*}
    J&=\alpha\int d^3\vec{r}d^3\mathbf{r}' \frac{\left|\phi(\mathbf{r})\right|^2\left|\phi(\mathbf{r}')\right|^2}{\left|\mathbf{r}-\mathbf{r}'\right|}=\alpha\int_0^\infty dr\left[\frac{1}{r} \left(1-e^{-\frac{2 r}{a}} \right)-\frac{1}{a}e^{-2r/a}\right]\frac{1}{\pi a^3}e^{-2r/a}4\pi r^2=\frac{5\alpha}{8a}
\end{align*}
and the Hartree-Fock equation reduces to
\begin{equation}
    -\frac{1}{2M}\nabla^2_{CM}\psi-\frac{1}{2\mu}\nabla^2\psi-\frac{4\alpha}{r}\left[1-\frac{e^{-r/a}}{2}\left(2+\frac{r}{a}\right)\right]\psi+\frac{5\alpha}{8a}\psi=E\psi\ ,
    \label{hf_to_sch_cm}
\end{equation}
with $M=2m_e$. This equation differs from the exact Schr\"odinger one in \eqref{hf_to_sch_cm}, and it certainly does not return, as self-consistent solution, the ansatz of Eq.~(\ref{ansatze+e-}).
Therefore, to fix the parameters of the calculation for charm and bottom quarks, it is best to use the {\it exact} (nonrelativistic) Schr\"odinger solution as opposed to the approximate Hartree-Fock one, which additionally takes the uncertainty due to the center of mass correction.

\section{Heavy-quark bound states}

\subsection{pNRQCD potential at LO and NLO}\label{sec:gluon}

In leading order (LO) in $\alpha_S = \frac{g_s^2}{4\pi}$, the one-gluon exchange in the nonrelativistic limit is represented by the potential
\begin{equation}\label{eq:gluongeneral}
    V^g_{q\bar{q}}(r) = -\frac{\alpha_s}{r}\  \ _{kl}\!\bra{\text{color}}T^a_{ki}T^a_{lj}  \ket{\text{color}}_{ij} = -C_F \frac{\alpha_s}{r}
\end{equation}
For a singlet quark-antiquark state~\cite{Cohen:2014vta} and 
$ \ket{\text{color}}_{1_C} = \frac{\delta_{ij}}{\sqrt{3}}\ket{q_iq_j} $
using $N_c = 3$ closure,
$
    T^a_{ki}T^a_{jl} = \frac{1}{2}\lb \delta_{kl}\delta_{ij} - \frac{1}{3}\delta_{ki}\delta_{lj} \rb
$,
the contraction of Eq.(\ref{eq:gluongeneral}) yields the color factor
\begin{equation}
    C_F = \frac{1}{\sqrt{3}}\frac{1}{\sqrt{3}} \sum_{i,j,k,l,a} \delta^{ij}T^{a}_{ki}T^{a}_{jl}\delta^{lk} = \frac{4}{3} =-\kappa_c
\end{equation}
so the gluon-exchange  $V^g_{q\bar{q}}(r)$ in Eq.~(\ref{eq:gluongeneral}) is attractive in color-singlet quarkonium.

 The Hartree-Fock program for charmonium and bottomonium is then identical to the one of positronium, except for the mass, a factor $4/3$ in the Hartree term and a change of the coupling constant $\alpha\rightarrow \alpha_s$.
s-diagrams are generally suppressed by $1/m^2$ in pNRQCD \cite{Brambilla:2015rqa}, so we won't consider them for heavy quark systems.

\begin{figure}
    \centering
    \includegraphics[scale = 0.45]{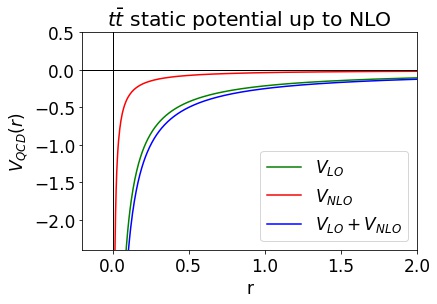}
    \caption{$t\bar{t}$ static potential up to NLO ($\alpha_s\simeq 0.16$).}
    \label{fig:vqcd}
\end{figure}

Instead, we consider the static potential to Next to Leading Order (NLO), in the same pNRQCD spirit, which is (with renormalization scale fixed to $r^{-1}$)
\begin{equation} \label{nlo}
    V_{NLO} (r) = -\frac{4}{3}\frac{\alpha_s^2}{4\pi r}\lb a_1 + 2\gamma_E\beta_0 \rb
\end{equation}
where $\gamma_E \approx 0.57721$ and 
\begin{equation}
    a_1 = \frac{31}{3} - \frac{10N_f}{9}\ ,\ \  \ \
    \beta_0 = 11 - \frac{2N_f}{3}
\end{equation}
$N_f$ being the number of flavors below the energy scale, $N_f=5$ at the $t$-quark scale, 4 and 3 at the $b$ and $c$ respectively. (An alternative would be to take $N_f=3$ in all three cases, after realizing that the energy scale $\frac{1}{2}m_q \alpha_s^2$ in pNRQCD is ultrasoft. The difference is small and well below other uncertainties.) The QCD static potential is illustrated in Figure \ref{fig:vqcd}.
Because $\alpha^2\equiv \alpha_{em}^2 \sim 10^{-4}$ is much smaller than $\alpha_s^2\sim 10^{-2}$, we do not correct the QCD static potential for electromagnetic interactions. However, in the next paragraphs we will examine Yukawa potentials due to other bosons from the electroweak sector because, while being weak, they bring a new functional dependence and serve as template for the new physics contact interactions to follow.
It remains to specify the coupling constant $\alpha_s$. As we are working at the scale of the charm quark and above~\cite{Deur:2016tte}, a perturbative treatment is reasonable and we adopt the Particle Data Group values and running~\cite{ParticleDataGroup:2024cfk}.

For two identical $q=c,b,t$ quarks, we choose  the scale as that quark's mass $Q^2=m_q^2$. 
For two different-flavoured quarks, in this work basically the mixed $cb$ interactions, we need a scale which interpolates between the 
two masses; due to the logarithmic running, it is convenient to use the geometric average 
$\alpha_s=\alpha_s\lb\sqrt{m_c^2 m_b^2}\rb$, a simple and sensible choice. To be slightly more precise, in the $t$-quark case we employ, in table~\ref{tab:top} below, $\alpha_{s}(m_t\alpha_s)$.

\subsection{Fitting procedure. Charmonium and bottomonium}
Here, we fit the charm and bottom masses to the weighted experimental values of the $\eta_c,\ J/\psi,\ \eta_b$ and $\Upsilon$ resonances, while running the coupling constant at NNLO in this section.
This is unlike the more self-contained procedure of~\cite{Mateu:2018zym} for the Cornell potential where $\alpha_s$ is also fit to the meson masses, we here adopt the standard value at the $Z$ pole and let it run down to the few GeV scale with the theory prediction.

Since the potentials used in this work are most often spin-independent, $J/\psi$ and $\Upsilon$ resonances will present spin degeneracy. Therefore, charmonium and bottomonium experimental masses~\cite{ParticleDataGroup:2024cfk} are weighted as
\begin{align}
    M_{c\Bar{c}}=\frac{M_{\eta_c}+3M_{J/\psi}}{4}&=3068.7\ \mathrm{MeV},\; \;  \mathrm{with} \; \; M_{\eta_c}=2983.9\ \mathrm{MeV}\ \mathrm{and}\ M_{J/\psi}=3096.9\ \mathrm{MeV}\nonumber\\
    M_{b\Bar{b}}=\frac{M_{\eta_b}+3M_{\Upsilon}}{4}&=9444.9\ \mathrm{MeV},\; \;  \mathrm{with} \; \; M_{\eta_b}=9398.7\ \mathrm{MeV}\ \mathrm{and}\ M_{\Upsilon}=9460.3\ \mathrm{MeV}\label{eq:experimental}
\end{align}
As in positronium, the exact ground state for $1s$ levels can be obtained from the Schrödinger equation, $E_{q\Bar{q}}(m_{q})=-\frac{1}{2}\mu(\frac{4}{3}\alpha_s(m_{q}))^2 $ where $\mu$ is the reduced mass, and the total mass is computed as $M_{q\Bar{q}} = 2m_q + E_{q\Bar{q}}$ at LO except for the running coupling $\alpha_s(m_q)$ values which are obtained at NNLO from the Z scale (but are considered the LO ones at $m_c$ and $m_b$) \cite{ParticleDataGroup:2024cfk}. Then the quark-mass parameter $m_q$ entering $\alpha_s(m_q)$ which minimizes the function $\chi^2(m_q)\equiv\left(M^{exp}_{q\Bar{q}}-M^{model}_{q\Bar{q}}(m_q)\right)^2$ is found for each meson.

These parameters serve in turn as inputs to the Hartree-Fock program. Hartree-Fock equations are identical to positronium's except for the mass and the $4/3$ color factor in the potential. The meson masses extracted from the program are compared with the ones of Schrödinger's and found to be compatible within the established precision. Moreover, we compute the mass of the $c\Bar{b}$ and its charge conjugate $b\Bar{c}$ and compare them with the $B_c^{\pm}$ experimental mass~\cite{ParticleDataGroup:2024cfk}. $B_c$ mesons are often studied in the bibliography to check the formalism used for charmonium and bottomonium calculations~\cite{S:2024wld}. Note that, however, differently-flavoured systems strongly suggest a different spatial orbital for each body. Therefore, each step in the SCF method requires to solve $N$-body Hartree-Fock equations. The results, presented in table \ref{tab:meson_fit}, are in good agreement with experimental data.
\begin{table}
  \centering
  \caption{Results for charmonium, bottomonium and $c\Bar{b}$   $1S$  meson  masses are presented in MeV for both the Schrödinger equation and Hartree-Fock calculations. These are compared with the experimental values~\cite{ParticleDataGroup:2024cfk}. Note that $m_{c\Bar{b}}=m_{b\Bar{c}}$. The parameters considered are $m_c=1575$ MeV, $m_b=4768$ MeV, from which  $\alpha_s^{c}=0.34$ and $\alpha_s^{b}=0.21$ follow.}
    \begin{tabular}{|c|ccc|}\hline
    \multicolumn{1}{|c|}{} & Schrödinger & Hartree-Fock & Experimental \\ \hline
    \hline
    $m_{c\Bar{c}}$ [MeV] & 3067  & 3078(18) & 3069 \\
    $m_{b\Bar{b}}$ [MeV] & 9440  & 9452(41) & 9445 \\
    $m_{c\Bar{b}}$ [MeV] & 6268  & 6272(18) & 6274 \\ \hline
    \end{tabular}%
  \label{tab:meson_fit}%
\end{table}%

We expect the approximation of taking the all-heavy system as Coulombic (the LO and NLO pNRQCD approximations) to continue holding some validity for the $p$-wave $\chi$ states, although with worse convergence and precision~\cite{Peset:2018jkf}. 

We therefore have a second set of parameters for which we also employ the spin-averaged mass of these
$1p$ states, according to 
\begin{equation}
    M_P=\frac{3M_{h_c}+M_{\chi_{c_0}}+3M_{\chi_{c_1}}+5M_{\chi_{c_2}}}{12}\ .
    \label{MP}
\end{equation}
This yields, for example, a second value for $m_c$ with the Schr\"odinger equation. If instead, the Hartree-Fock equation is 
employed, a slightly different value (at LO or at NLO) will be obtained. All the extractions are shown in table~\ref{fitparameters}.

\begin{table}[h]
\centering
\caption{Parameters obtained at LO and NLO in the $\alpha_s$ counting, in the nonrelativistic philosophy of pNRQCD. 
The spread in  $\alpha_s$ values is very small, and that in mass values, very reasonable. 
For part of the calculations we will adopt the computation in the last column with the uncertainty from table~\ref{parameters}.}
\label{fitparameters}
\begin{tabular}{c|c|c|c|c|c|c|}
\cline{2-5} \cline{7-7}
   & Schrödinger &Schrödinger & HF & HF &  & 
   \begin{tabular}[c]{@{}c@{}}Reference \end{tabular} \\
   & $p$ wave &$s$ wave & LO & NLO &  & 
   \begin{tabular}[c]{@{}c@{}}values \end{tabular} 
\\ \cline{1-5} \cline{7-7} 
\multicolumn{1}{|c|}{$m_c$ (MeV)}       & 1773                                         & 1578   & 1568 & 1607                           &  & 1590                                                                                    \\ \cline{1-5} \cline{7-7} 
\multicolumn{1}{|c|}{$\alpha_s(m_c^2)$} & 0.320                                        & 0.338                                        & 0.334                         & 0.332                          &  & 0.333                                                                                   \\ \cline{1-5} \cline{7-7} 
\multicolumn{1}{|c|}{$m_b$ (MeV)}       & 4962                                         & 4771                                         & 4765                          & 4791                           &  & 4780                                                                                    \\ \cline{1-5} \cline{7-7} 
\multicolumn{1}{|c|}{$\alpha_s(m_b^2)$} & 0.214                                        & 0.216                                        & 0.215                         & 0.215                          &  & 0.215                                                                                   \\ \cline{1-5} \cline{7-7} 
\end{tabular}
\end{table}

The differences among the extractions of  $m_b$, $m_c$, $\alpha_s(m_b^2)$ and $\alpha_s(m_c^2)$ provides us with a systematic uncertainty to be propagated to successive predictions. Incidentally, as there could be correlated errors we will be conservative
and use linear uncertainty propagation  $|\Delta|=\sum_i|\Delta_i|$, see table \ref{parameters}, instead of the uncorrelated root-mean square. 
Given the small uncertainty assigned to the coupling constant, this means that the quark masses's uncertainty propagates as
\begin{equation}
    \Delta M\leq \sum_i\Delta m_i+\sum_q\bigg|\frac{\partial E}{\partial m_q}\bigg|\Delta m_q
    \label{uncertainties}
\end{equation}
with the first sum running over all particles whose masses are added up to form the composite mass, and the second over $m_c$ and $m_b$, giving the indirect uncertainty through the binding energy. The derivative is computed by finite differences, and the resulting $\Delta M$ values are also shown in table~ \ref{ccbbresult}.

\begin{table}[h]
\centering
\caption{Second computation of $c\bar{c}$, $b\bar{b}$ and mixed-flavor $B_C$ and experimental reference values.}
\label{ccbbresult}
\begin{tabular}{c|cc|cc|cc|}
\cline{2-7} & \multicolumn{2}{c|}{$c\bar{c}$} & \multicolumn{2}{c|} {$b\bar{b}$} & \multicolumn{2}{c|}{$B_c$} \\ 
\cline{2-7} & \multicolumn{1}{c|}{S} & P & \multicolumn{1}{c|}{S} & P & \multicolumn{1}{c|}{S} & P \\ \hline \multicolumn{1}{|c|}{LO}  & \multicolumn{1}{c|}{3100(80)} & 3160(80)& \multicolumn{1}{c|}{9470(60)} & 9540(60)  & \multicolumn{1}{c|}{6310(70)} & 6370(70)  \\ \hline
\multicolumn{1}{|c|}{NLO}  & \multicolumn{1}{c|}{3030(80)} & 3140(80) & \multicolumn{1}{c|}{9410(60)}   & 9520(60)  & \multicolumn{1}{c|}{6270(70)}  & 6350(70) \\ \hline
\multicolumn{1}{|c|}{\begin{tabular}[c]{@{}c@{}}Expt. spin\\ average \cite{ParticleDataGroup:2022pth} \end{tabular}} & \multicolumn{1}{c|}{3069}  & 3526  & \multicolumn{1}{c|}{9445}  & 9889  & \multicolumn{1}{c|}{6274}     & 6714   \\ \hline
\end{tabular}
\end{table}

Globally, there is a reasonable compatibility with the experimental spin averages. The $s$-wave states are predicted to 1\% precision.
The   $p$-wave shows larger deviations,  consistently with the difference $M_P-M_S$ in table \ref{parameters} and points out that the $p$-wave state are less deep in the Coulomb regime and start perceiving higher-order terms in the potential or even the nonperturbative linear piece.  To this uncertainty we should also add the imprecise CM correction.
This means that the fit parameters suffer from a systematic error which will be propagated to later computations. For the rest of the work we will limit ourselves to  $s$-wave states, so we do not need the fit including the $p$-wave quarkonium, but we show it to clearly expose some  of the limitations of the current approach.

It happens that the LO potential slightly overshoots, and the NLO potential undershoots (although by much less, as expected) the data.

\begin{table}[h]
\centering
\caption{Final parameter set for all-$c,b$ multiquark hadrons.}
\label{tab:finalparameters}
\begin{tabular}{|c|c|c|c|}
\hline
$m_c$ (MeV) & $\alpha_s(m_c^2)$ & $m_b$ (MeV) & $\alpha_s(m_b^2)$ 
\\ \hline
1575  & 0.34  & 4768 & 0.21   \\ \hline
\end{tabular}
\end{table}

\begin{table}[h]
\centering
\caption{Sizeable uncertainty sources to be propagated.}
\label{parameters}
\begin{tabular}{|c|cccc|c|l|c|}
\cline{1-6} \cline{8-8}
& \multicolumn{4}{c|}{Difference to employed average}  &CM &  &  \\ 
& $p$-wave &  $s$-wave & LO & NLO & recoil    & & Uncertainty  \\ 
\cline{1-6} \cline{8-8} 
\multicolumn{1}{|c|}{$m_c$ (MeV)}       & \multicolumn{1}{c|}{180}                              & \multicolumn{1}{c|}{10}                               & \multicolumn{1}{c|}{20}                         & 20                          & 20                            &  & 40                                                      \\ \cline{1-6} \cline{8-8} 
\multicolumn{1}{|c|}{$\alpha_s(m_c^2)$} & \multicolumn{1}{c|}{0.013}                            & \multicolumn{1}{c|}{0.005}                            & \multicolumn{1}{c|}{0.001}                      & 0.001                       & 0.005                         &  & 0.006                                                   \\ \cline{1-6} \cline{8-8} 
\multicolumn{1}{|c|}{$m_b$ (MeV)}       & \multicolumn{1}{c|}{180}                              & \multicolumn{1}{c|}{10}                               & \multicolumn{1}{c|}{10}                         & 10                          & 20                            &  & 30                                                      \\ \cline{1-6} \cline{8-8} 
\multicolumn{1}{|c|}{$\alpha_s(m_b^2)$} & \multicolumn{1}{c|}{0.001}                            & \multicolumn{1}{c|}{0.001}                            & \multicolumn{1}{c|}{0.001}                      & 0.001                       & 0.003                         &  & 0.004                                                   \\ \cline{1-6} \cline{8-8} 
\end{tabular}
\end{table}

\section{Spin-averaged all-$c$ and $b$ spectra I:  tetraquarks}\label{section:results1}
In this section we present the ground-state wavefunctions and  tetraquark spectra.  We will compare the results with several of the many prior calculations~\cite{Lundhammar:2020xvw,Debastiani:2017msn,Lin:2024olg,Wu:2016vtq,Wang:2018poa,Bedolla:2019zwg}.
We will treat the system in two variational approximations: as a four-body Hartree-Fock problem and as an iterated two-body problem based on the diquark-antidiquark approach.

\subsection{Tetraquark color structure}\label{subsection:tetraquarkcolorstructure}

We will here stay clear of the delicate large-$N_c$ tetraquark~\cite{Cohen:2014vta,Lucha:2017gqq}
colour structure and set $N_c=3$.
We first examine the compact tetraquark configuration formed by a diquark-antidiquark pair~\cite{Lundhammar:2020xvw,Debastiani:2017msn,Drenska:2008gr}, as opposed to a meson-meson molecule. These diquarks are not in a singlet state, and hence the resulting tetraquark system is expected to be a strongly bound state due to gluon exchange.   The color wavefunction for diquarks (antidiquarks) in the antitriplet (triplet) state is $\ket{\xi_{qq}}=\frac{1}{\sqrt{6}} \varepsilon^{\alpha ij}c_i c_j$, where $\varepsilon^{\alpha i j}$ connects the two incoming quarks so that the combination has zero net color charge \cite{cbb_barcelona}. 
Now the color factor  is
\begin{equation}
    f_{qq} = \mel{\xi_{qq}}{T^aT^a}{\xi_{qq}}= \frac{{\varepsilon_\alpha}^{kl}}{\sqrt{6}}\left(c_k T^a c_i\right)\left(c_l T^a c_j\right)\frac{\varepsilon^{\alpha ij}}{\sqrt{6}}=-\frac{2}{3}.\label{eq:qqcolorfactor}
\end{equation}
Contrary to electrodynamics, the interaction potential for gluon exchange within a diquark $-\frac{2}{3}\frac{\alpha_s}{r}$ is atractive. From the point of view of colour, this system is akin to the gluon exchange between two quarks within a baryon (which is in a singlet state). In this case, the \emph{dummy} index $\alpha$ of Eq.~(\ref{eq:qqcolorfactor}) turns out to be the color state of the non-interacting quark.

The $u$-channel diagram where the final state quarks are interchanged, introducing the corresponding statistical factor, happens to be the same as in Eq.~(\ref{eq:qqcolorfactor}) \cite{Viniciusthesis}.

The colour wavefunction of two quarks and two antiquarks can  be decomposed as
\begin{equation}
\ket{qq\Bar{q}\Bar{q}}:3\otimes 3\otimes \Bar{3}\otimes \Bar{3}=(3\otimes3)\otimes(\Bar{3}\otimes\Bar{3})=\Bar{3}\otimes3 + 6\otimes\Bar{6}+\Bar{3}\otimes\Bar{6}+6\otimes3,
\end{equation}
where the color singlet states correspond to $\Bar{3}\otimes3$ (antisymmetric under $1\leftrightarrow2$ and $3\leftrightarrow4$) and $6\otimes\Bar{6}$ (symmetric). In the notation of \cite{Buccella:2006fn}, these can be read as $(q_1q_2)^{\Bar{3}}\otimes(\Bar{q}_3 \Bar{q}_4)^3 \; \mathrm{and}\; (q_1 q_2)^6\otimes(\Bar{q}_3 \Bar{q}_4)^{\Bar{6}}$ respectively. With this notation it is easy to interpret the tetraquark as a meson composed by a diquark and an antidiquark. Therefore, its spectrum can be computed either in the diquark-antidiquark picture or as a four-body problem \cite{Debastiani:2017msn,Lin:2024olg}. For the latter we must calculate the relevant color factors.

A tetraquark color-singlet basis  can then be taken~\cite{Park:2013fda} as 
\begin{align}
    (q_1q_2)^{\Bar{3}}\otimes (\Bar{q}_3\Bar{q}_4)^3 &= \frac{1}{\sqrt{12}}\varepsilon^{\alpha ij}\varepsilon_{\alpha kl}c_i(1)c_j(2)c^{\dagger k}(3)c^{\dagger l}(4)\\
    (q_1 q_2)^6\otimes(\Bar{q}_3 \Bar{q}_4)^{\Bar{6}} &= \frac{1}{\sqrt{6}}d^{\alpha i j}d_{\alpha kl} c_i(1)c_j(2)c^{\dagger k}(3)c^{\dagger l}(4)
\end{align}
where $\epsilon^{ijk}$ is the Levi-Civita antisymmetric and $d^{\alpha ij}=d_{\alpha ij}$ the symmetric symbols
\begin{equation}
    d_{111}=d_{222}=d_{333}=1,\; \; d_{412}=d_{421}=d_{523}=d_{532}=d_{613}=d_{631}=\frac{1}{\sqrt{2}},
\end{equation}
the rest of them 0. Note that these wave functions are already normalized. 

For the interaction between the i-th and j-th quarks, the color factor for a tetraquark in the $\Bar{3}\otimes3$ state can be obtained as follows
\begin{equation}
    f^T_{qq} = \frac{1}{12}\varepsilon^{\beta mn}\varepsilon_{\beta pq} T^a_{mi}T^a_{nj}\varepsilon^{\alpha ij}\varepsilon_{\alpha kl}\delta^{pk}\delta^{ql} = -\frac{2}{3}=\kappa_{c,qq}^T
\end{equation}
where $\delta^{pk}\delta^{ql}$ arise from the fact that color must be conserved for the non-interacting antiquarks. In the same way, the interaction between the i-th quark and the k-th or l-th antiquark yields a color factor
\begin{equation}
    f^T_{q\Bar{q}} = \frac{1}{12}\varepsilon^{\beta mn}\varepsilon_{\beta pq} T^b_{mi}T^b_{kp}\varepsilon^{\alpha ij}\varepsilon_{\alpha kl}\delta^{nj}\delta^{ql} = \frac{1}{3}=-\kappa_{c,q\Bar{q}}^T
\end{equation}
The results for the $6\otimes\Bar{6}$ state are analogously obtained, yielding $f^S_{qq}=\frac{1}{3}$ and $f^S_{q\Bar{q}}=\frac{5}{6}$. Hence, diquarks in the sextet state see a repulsive potential and should not bind \cite{Lundhammar:2020xvw}. Detailed tetraquark calculations \cite{Debastiani:2017msn} in the diquark-antiquark picture suggest that the tetraquark is not found in the $6\otimes\Bar{6}$ state. Finally, in \cite{Lin:2024olg,Park:2013fda,Wu:2016vtq,Liu:2019zuc} tetraquark calculations in the four-body picture evaluate the mixing of sextet diquarks with antitriplet ones.

\subsection{Tetraquark spin structure}
The tetraquark spin structure can be easily understood in the diquark-antidiquark picture \cite{Maiani:2014aja,Maiani:2004vq}. Since the color wave function of the constituent diquarks is taken to be in the antisymmetric antitriplet state, the spin part must be symmetric in order to fulfill fermion antisymmetry, i.e. $S_{qq}=S_{\Bar{q}\Bar{q}}=1$. The computed diquarks will therefore be in the $N^{2s+1}L_J=1^3S_1$ configuration. Coupling diquark and antidiquark yields a tetraquark spin  $S_T=0,1,2$. Therefore, the s-wave $(L_T=0)$ symmetrically-flavoured tetraquark can be in the $J^{PC}=0^{++},\ 1^{+-}$ and $2^{++}$ states, where $J$ is the total angular momentum of the tetraquark, $P=(-1)^{L_T}$ is the parity and $C=(-1)^{L_T+S_T}$ is the charge conjugation. Since the diquark and antidiquark couple similarly to mesons, we can reuse some well known parity and charge conjugation rules \cite{Lundhammar:2020xvw,Viniciusthesis}. 
Of course, for spin-independent potentials, these three states will degenerate.

Since tetraquarks in the $\Bar{3}\otimes3$ colour state are composed of quarks in different colour-spin orbitals, all Fock terms cancel out. Thus, considering the color factors presented in section \ref{subsection:tetraquarkcolorstructure}, the Hartree-Fock equations for a quark within the S-wave same-flavoured tetraquark can be derived from Eq.~(\ref{eq:Hartree_Fock_QCD}), 
\begin{equation}
	\mathcal{F}\psi = -\frac{3}{8m}\nabla^2\psi(q) - \frac{4}{3}\int{d^3\mathbf{r}'\left|\phi(\mathbf{r}')\right|^2\frac{\alpha_s}{\left|\mathbf{r}-\mathbf{r'}\right|}\psi(q)},
 \label{eq:fock_tetraquark}
\end{equation}
where we have considered $\psi_i=\psi_j=\psi_k=\psi_l$ and thereby the factor $-\frac{2}{3}$ from the quark-quark interaction and $2\cdot\left(-\frac{1}{3}\right)$ from the quark-antiquark interactions are summed. The Hartree-Fock equations are identical for the other three bodies all being in the same $1s$ orbital. We lift the orbital degeneracy when computing differently-flavoured tetraquarks, for which we solve two coupled equations (one for $c$ and a second for $b$) in each SCF step. The binding energy is obtained from Eq.(\ref{energy}).

\subsubsection{Diquark-antidiquark setup}
Our second calculation to have a comparison point, within the diquark-antidiquark approach requires to compute the diquark wavefunction first. Diquarks, as colored objects, are absent from the physical spectrum. However, they are useful auxiliary quantities inside a hadron, just as quarks are, and like these, diquarks are also assigned a mass parameter which must be understood as a short-distance quantity. Assuming diquarks with flavour-symmetric wavefunctions and antiparallel spins in the $1^3S_1$ configuration, the Hartree-Fock equations for a diquark constituent are identical to the meson's except for the color factor, which in this case is $-2/3$. Once the diquark masses are computed, the Hartree-Fock equations for the diquark-antidiquark system are solved
\begin{equation}
	\mathcal{F}\psi = -\frac{1}{4m_{diquark}}\nabla^2\psi(q) - \frac{4}{3}\int{d^3\mathbf{r}\left|\phi(\mathbf{r}')\right|^2\frac{\alpha_s}{\left|\mathbf{r}-\mathbf{r'}\right|}\psi(q)}.
 \label{eq:fock_tetraquark_diquarks}
\end{equation}
Notice that the color factor in this case is $4/3$, since the diquark-antidiquark system is in a $\Bar{3}\otimes3$ state, and therefore its color factor should correspond to the same color singlet as the meson's. Diquark results are obtained from both the Schrödinger equation $E_{qq}(m_{q})=-\frac{1}{2}\mu(\frac{2}{3}\alpha_s(m_{q}))^2$ and the Hartree-Fock program. These are presented and compared with a few selected earlier works in table \ref{tab:diquarks}.
\begin{table}
  \centering
  \caption{Computed masses for auxiliary $1^3S_1$ heavy diquarks are presented (in MeV) and compared to previous work. The parameters are those fit to the conventional meson spectrum (table \ref{tab:meson_fit}). This is not a physical hadron mass, but is to be understood as a short-distance quantity useful within a perturbative approach to pNRQCD, losing usefulness whenever the confining potential is probed.}
    \begin{tabular}{|c|cc|ccccc|} \hline
          & Schrödinger & Hartree-Fock & Ref. \cite{Lundhammar:2020xvw} & Ref. \cite{Debastiani:2017msn} & Ref. \cite{Galkin:2023wox} & Ref. \cite{Kiselev:2002iy} & Ref. \cite{Bedolla:2019zwg}\\
    \hline
    $m_{cc}$ & 3130  & 3132(7) & 3128  & 3133  & 3226  & 3130 & 3329\\
    $m_{bc}$ & 6324  & 6326(8) & --    & --    & 6526  & 6450 & 6611\\
    $m_{bb}$ & 9513  & 9516(10) & 9643  & --    & 9778  & 9720 & 9845\\ \hline
    \end{tabular}%
  \label{tab:diquarks}%
\end{table}%

First, the results of the direct HF tetraquark computation and the one passing by an auxiliary diquark-antidiquark pair are identical for all purposes.

Second, the computed auxiliary diquark masses for the $c$ quark are found to be in reasonable agreement with the bibliography, which presents some spread.

Third, our computed diquarks involving the $b$ are somewhat lighter than those reported by other groups.  This is not necessarily a concern, since $m_{qq}$ is an auxiliary parameter and, as we next show, the physical all-heavy tetraquark masses are in line with earlier works. 

\subsubsection{All-heavy Tetraquark spectrum}
Computed masses for $cc\Bar{c}\Bar{c}$ and $bb\Bar{b}\Bar{b}$ are presented in table \ref{tab:tetraquarks} for both the four-body problem and the diquark-antidiquark picture. Remember that we have considered correlated errors, conservatively adding them linearly and not in quadrature so that $\Delta M_{qq\Bar{q}\Bar{q}}=2\Delta E_{qq}+\Delta E_{qq\Bar{q}\Bar{q}}$. Importantly, we run again the coupling constant for tetraquark calculations in the diquark-antidiquark picture to the scale of the diquark masses $m_{qq}$, since they are significantly different than the ones of $m_c$ or $m_b$.

\begin{table}
  \centering
  \caption{Comparison of computed masses for heavy $1s$-orbital tetraquark in HF, treated in both the diquark-antidiquark picture (1st column) and as a four-body problem (2nd column), with results from other works. Since we are calculating spin-degenerate states, we have performed a weighted average of the $0^{++},1^{+-}$ and $2^{++}$ masses from earlier works, in line with Eq.(\ref{eq:experimental}). Employed quark masses  are those of table \ref{tab:meson_fit}, $m_{cc}=3130$, $m_{bc}=6324$, and $m_{bb}=9513$ MeV, respectively. The scale choice for the coupling constant yields $\alpha_s(m_{cc})=0.24$, $\alpha_s(m_{bb})=0.18$ and $\alpha_s(m_{bc})=0.20$.}
  \begin{adjustbox}{max width=\textwidth}
    \begin{tabular}{|c|cc|ccccccc|} \hline 
          & $qq-\Bar{q}\Bar{q}$ & $qq\Bar{q}\Bar{q}$ (4-body) & Ref. \cite{Lundhammar:2020xvw} & Ref. \cite{Lundhammar:2020xvw} & Ref. \cite{Debastiani:2017msn} & Ref. \cite{Wang:2018poa} & Ref. \cite{Bedolla:2019zwg} & Ref. \cite{Wu:2016vtq} & Ref. \cite{Lin:2024olg} \\ \hline
    $m_{cc\Bar{cc}}$ [MeV] & 6193(32) & 6205(13) & 6054  & 6283  & 6068  & 6066(80) & 6164  & 6919  & 6300 \\
    $m_{cc\Bar{bb}}$ [MeV] & 12562(39) & 12589(13) & --    & --    & --    & --    & 12569 & 13572 & -- \\
    $m_{bc\Bar{bc}}$ [MeV] & 12553(39) & 12590(13) & --    & --    & --    & --    & 12525 & 13531 & -- \\
    $m_{bb\Bar{bb}}$ [MeV] & 18912(50) & 18963(14) & 18753 & 18783 & --    & 18850(90) & 18859 & 20223 & -- \\ \hline
    \end{tabular}%
    \end{adjustbox}
  \label{tab:tetraquarks}%
\end{table}%
The masses obtained for $cc\Bar{c}\Bar{c}$, $cc\Bar{b}\Bar{b}$ and $bc\Bar{b}\Bar{c}$ tetraquarks are consistent with some other works, making allowance for the large dispersion of earlier computations. Our results are in specially good agreement with Ref.\cite{Bedolla:2019zwg}, which uses a relativized diquark model in the antitriplet-triplet color configuration. 

In spite of the uncertainty due to the CM energy correction to the two-body problem, the masses computed from the diquark-antidiquark wavefunction do not look significantly different from those in the direct tetraquark computation (with larger $1/N$ suppression factor for this correction), so it seems to be under control.

\begin{SCfigure}[0.55][t]
    \caption{Ground state $cc\Bar{c}\Bar{c}$ tetraquark mass in the diquark-antidiquark picture and in the 4-body problem. The meson-meson decay thresholds are shown, together with  X(6200) and X(6600) structures from the ATLAS and CMS collaborations \cite{CMS:2023owd,ATLAS:2023bft}.}
\includegraphics[width=.7\textwidth]{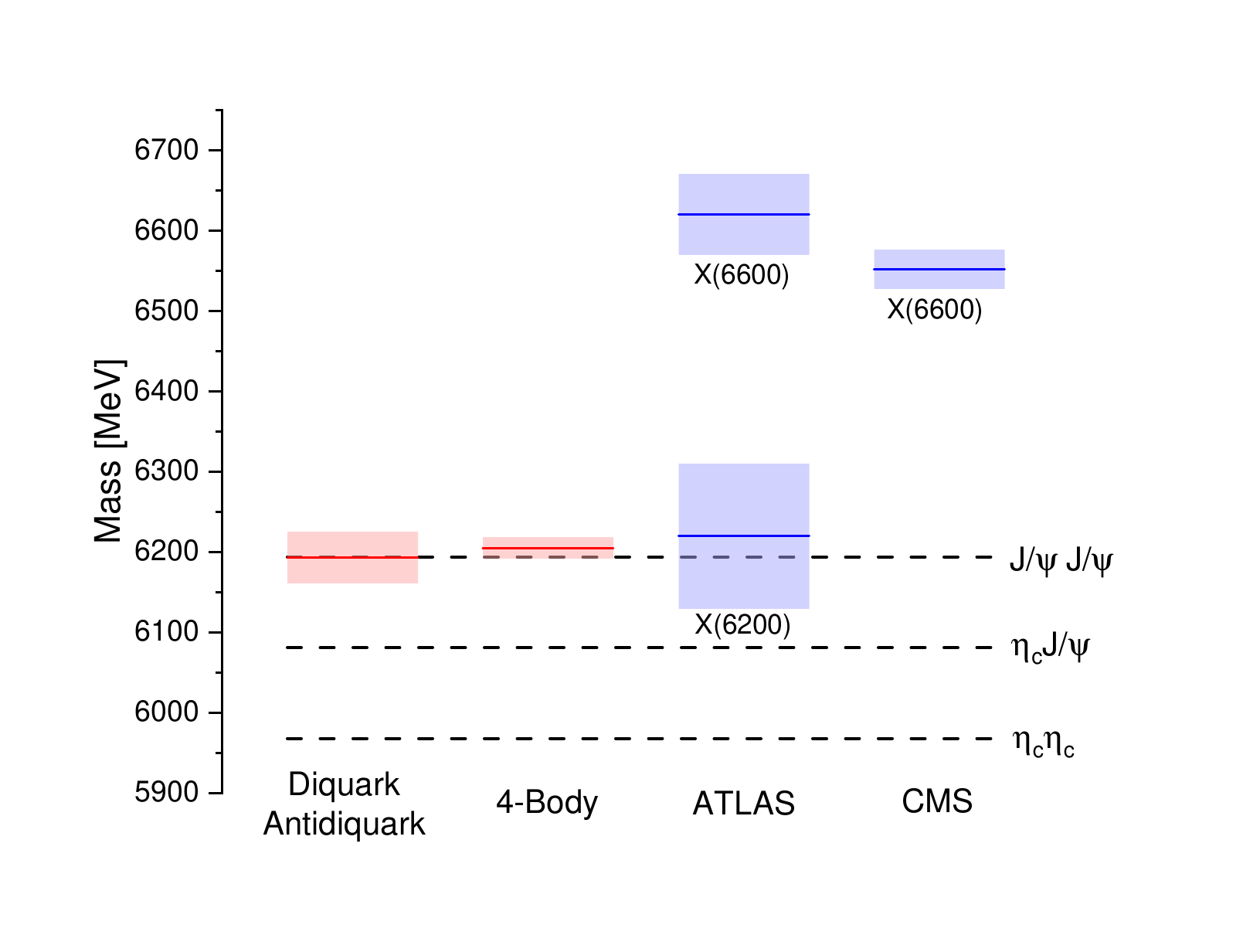}
\label{fig:grotriancccc}
\end{SCfigure}

Fig.~\ref{fig:grotriancccc} is a Grotrian diagram for the $cc\Bar{c}\Bar{c}$ tetraquarks,  marked with open thresholds for meson pairs to which decays are possible, and experimental structures in the $J/\psi J/\psi$ channel. The $s$-wave tetraquarks have $J^{PC}=0^{++},1^{+-}$ and $2^{++}$ as degenerate states. The ground states of three possible meson combinations are $\eta_c\eta_c: 0^{++},\ \eta_c J/\psi:1^{+-}$ and $J/\psi J/\psi:0^{++},1^{++},2^{++}$. Since the computed tetraquark masses lie very close to the $J/\psi J/\psi$ threshold, within our uncertainties  we can not state whether they can decay into two $J/\psi$ mesons. However, they can surely decay into the pairs involving the lighter $\eta_c$.

The same can be said for the $bb\Bar{b}\Bar{b}$ tetraquark. The $bc\Bar{b}\Bar{c}$ and $bb\Bar{c}\Bar{c}$ tetraquarks lie close to their respective $B_c^{+}B_c^{-}$ and $J/\psi\Upsilon$ thresholds as well, but can decay into $\eta_c\eta_b,\ \eta_c\Upsilon$ and $J/\psi\ \eta_b$ mesons. There is no experimental evidence yet for any assignable resonances.

The X(6200) structure detected by the ATLAS collaboration \cite{ATLAS:2023bft} might be  compatible with a resonance including an admixture of a compact component in its wave function \cite{Nefediev:2021pww}, and our $cc\Bar{c}\Bar{c}$ tetraquark calculation  passes the sanity test yielding a similar mass, although a molecular interpretation is also compatible with the experimental mass \cite{Dong:2020nwy}.

\section{Spin-averaged all-$c$ and $b$ spectra II: baryons, pentaquarks and dibaryons.}\label{section:results2}
\subsection{All-heavy baryons and pentaquarks}
Turning to baryons, no three-heavy-quark ones have yet been found, but the detection of doubly-charmed ones at the LHC makes the prospect ever closer. This means that, as for tetraquarks, no experimental data can anchor the calculations yet. The $\Omega_{ccc}$ and $\Omega_{bbb}$ baryons must, by force of fermion antisymmetry, have spin-parity $\frac{3}{2}^+$ in their ground state. It suffices (given rotational invariance) to compute the mass of the $m_J=3/2$ state 
\begin{equation}
    \Omega_{ccc}=\chi_+\chi_+\chi_+\otimes ccc\ ,
    \label{cccstate}
\end{equation}
and equivalently for the $b$ quark. 
Since, excepting colour, the rest of the wavefunction is symmetric, the Fock terms end up adding to the Hartree ones with the same sign.
The Hartree-Fock equations~(\ref{altpot}) then become \footnote{All equations in this subsection are understood to carry the factor  $(1-1/N)$ in the kinetic-energy term to (at least partially) correct for the CM recoil.}
\begin{equation}
    -\frac{1}{2m_c}\frac{d^2P}{dr^2}+4\mathcal{V}P=\varepsilon P.
    \label{ccc}
\end{equation}
After solving this eigenvalue problem,ying Koopman's theorem yields the masses in table~\ref{baryon1results}.

The mixed-flavor ones, $\Omega_{ccb}$ and $\Omega_{bbc}$, can present themselves (as do light baryons $N$ and $\Delta$) with spins   $(1/2)^+$ and  $(3/2)^+$. Upon ignoring spin interactions, it could be expected that  we predict the mass average only
\begin{equation*}
    M=\frac{2M_{1/2}+4M_{3/2}}{6}=\frac{M_{1/2}+2M_{3/2}}{3}\ .
\end{equation*}
And yet, as the two wavefunctions are distinct due to symmetry, a spin splitting arises. 
Their completely symmetric spin-flavour wavefunctions are then
\begin{eqnarray}
    \Omega_{ccb} =\frac{1}{\sqrt{2}}\left(\chi_S\otimes\Omega_S+\chi_A\otimes\Omega_A\right) =\frac{1}{\sqrt{18}}(
    ccb\otimes(2\chi_+\chi_+\chi_- - \chi_+\chi_-\chi_+ - \chi_-\chi_+\chi_+)
    +\nonumber\\
    +cbc\otimes(2\chi_+\chi_-\chi_+ - \chi_-\chi_+\chi_+ - \chi_+\chi_+\chi_-)
    +\nonumber +bcc\otimes(2\chi_-\chi_+\chi_+ - \chi_+\chi_-\chi_+ - \chi_+\chi_+\chi_-))\ .
    \label{ccb12state}
\end{eqnarray}
Instead, for  $\Omega_{ccb}$ with spin $3/2$, the wavefunction becomes
\begin{align}
    \Omega_{ccb}&=\frac{1}{\sqrt{3}}\left(ccb+cbc+bcc\right)\otimes\chi_+\chi_+\chi_+.
    \label{ccb32state}
\end{align}

For  $\Omega_{ccb}$-like baryons, with quarks of two different masses, we will naturally augment the variational function space by allowing different radial orbitals $P_c$ and $P_b$, which will solve coupled equations; their spin-flavor part is as discussed in Eq.~(\ref{ccb12state}) and \eqref{ccb32state}. 
For example, for $ccb\otimes\chi_+\chi_-\chi_+$,  substituting in Eq.~(\ref{altpot}) yields
\begin{align}
    \left \{
    \begin{array}{cc}
         &  -\frac{1}{2m_c}\frac{d^2P_c}{r^2}+(\mathcal{V}_{cc}+\mathcal{V}_{bb})P_c=\varepsilon_cP_c\\
         & -\frac{1}{2m_b}\frac{d^2P_b}{dr^2}+2\mathcal{V}_{cc}P_b=\varepsilon_bP_b
    \end{array}
    \right. \ . 
    \label{ccb}
\end{align}
The total energy will then be obtained taking account of the coefficients in Eq.~(\ref{ccb12state}) and~(\ref{ccb32state}).
The numerical values of the obtained energies are then carried onto table~\ref{baryon1results}, together with their uncertainties from~\eqref{uncertainties}.

\begin{figure}
\centering
\includegraphics[width=0.5\textwidth]{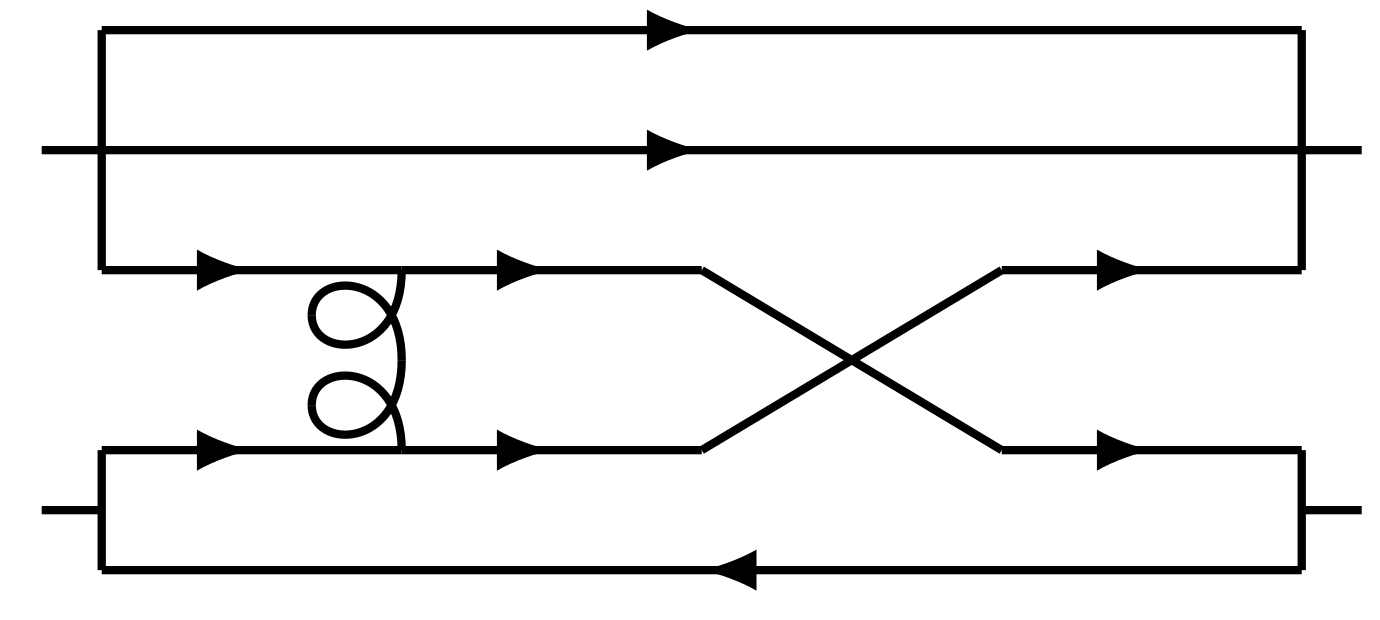}
\caption{Colour interactions among two colour-singlet clusters such as baryon-meson to form a pentaquark: gluon exchange must be accompanied by quark exchange to balance colour. 
\label{fig:pentaFeynman}}
\end{figure}

To address pentaquarks we will make use of their easy decomposition into a baryon and a meson colour-singlet clusters.  The colour factors inside each cluster are those already at hand, and the only additional one (remember that $s$-channel annihilation is suppressed by  $\Lambda_{\rm QCD}/m$) is the $u$-channel exchange depicted in figure~\ref{fig:pentaFeynman} and typical of the Resonating Group Method~\cite{Bicudo:1987tz,Entem:2000mq}.

Unlike in three-quark baryons, where $u$-channel diagrams are equivalent to Fock terms in Hartree-Fock theory, here this diagram needs separate consideration.

The corresponding colour factor is then
\begin{equation}
    F_c=\left(\frac{1}{\sqrt{6(3+1)}}\epsilon_{ijk}\delta_{pq}\right)\left(T_{kn}^aT_{pr}^a\delta_{il}\delta_{jm}\delta_{qs}\right)\left(\frac{1}{\sqrt{6(3+1)}}\epsilon_{lmn}\delta_{rs}\right)=\frac{1}{3}.
\end{equation}

\subsection{Spin-dependent potential: baryons as an example}
Although most of the work presented in this article, for heavy particles, employs spin-independent interactions, we make an exception in this section to assess the spin splittings.
The all-heavy baryon masses will be recomputed with the standard hyperfine interaction~\cite{Thomas:2001kw}
\begin{equation}
    V=F_c\left(-\frac{\alpha_s}{r}+\frac{8\pi}{3}\frac{\alpha_s}{m_i m_j}\delta^3(\mathbf{r})\mathbf{S}_i\cdot\mathbf{S}_j\right),
    \label{Vspin}
\end{equation}
with $m_i$ and $m_j$ the masses of the quarks exchanging the gluon, to see the typical effect (without attempting a full calculation with all other $O(m^{-2})$ effects included). 

This adds a Dirac delta to the earlier Coulomb integral  of $\mathcal{V}$.
The new contribution to the HF equation (which will again appear when we handle the $t$-quark later)
\begin{equation*}
    \int \delta(r-r')R_i(r')R_j(r')dr'=\frac{P_i(r)P_j(r)}{r^2},
\end{equation*}
multiplied by the prefactors in \eqref{Vspin}. As usual, the spin matrix element is exchanged (for $s_i=1/2$) for the total spin,
\begin{equation}
    S_{ij}^2=(\mathbf{S}_i+\mathbf{S}_j)^2=S_i^2+S_j^2+2\mathbf{S}_i\cdot\mathbf{S}_j=\frac{3}{2}+2\mathbf{S}_i\cdot\mathbf{S}_j\ .
\end{equation}

\begin{table}
\centering
\caption{Baryon masses (in MeV) for two central potentials (LO and NLO) and for the latter including, additionally, the spin-spin hyperfine contact interaction. No experimental data is yet available; for a comparison with some other theoretical predictions see our previous publications on these baryons~\cite{Llanes-Estrada:2011gwu,deArenaza:2024dhe}. The benchmark quark-model work of Silvestre Brac~\cite{Silvestre-Brac:1996myf} that encompassed many earlier calculations quotes masses about 100 MeV systematically above these computations.}\label{baryon1results}
\begin{tabular}{c|c|c|c|}
\cline{2-4}  &  LO     &  NLO    &  NLO+Spin   \\ \hline
\multicolumn{1}{|c|}{ {  $\Omega_{ccc}\left(\frac{3}{2}\right)^+$}} & {  4650(120)} & {  4660(120)} & {  4660(120)} \\ \hline
\multicolumn{1}{|c|}{ {  $\Omega_{ccb}\left(\frac{1}{2}\right)^+$}} & 7920(110)                        & 7930(110)                        & 7920(110)                        \\ \hline
\multicolumn{1}{|c|}{ $\Omega_{ccb}\left(\frac{3}{2}\right)^+$}                        & 7920(110)                        & 7920(110)                        & 7920(110)                        \\ \hline
\multicolumn{1}{|c|}{ {  $\Omega_{bbc}\left(\frac{1}{2}\right)^+$}} & 11100(100)                       & 11080(100)                       & 11100(100)                       \\ \hline
\multicolumn{1}{|c|}{ $\Omega_{bbc}\left(\frac{3}{2}\right)^+$}                        & 11100(100)                       & 11090(100)                       & 11100(100)                       \\ \hline
\multicolumn{1}{|c|}{ $\Omega_{bbb}\left(\frac{3}{2}\right)^+$}                        & 14280(90)                       & 14270(90)                       & 14280(90)                       \\ \hline
\end{tabular}
\end{table}

Without fatiguing the reader with similar detail, we immediately quote our computed masses for all-heavy pentaquark baryons in table~\ref{pentaquarktable}, limiting ourselves to the case of $s$-wave orbitals, employing the meson-baryon basis and introducing both the NLO and the spin-spin potential in the same terms and  parameters as for ordinary baryons. See~\cite{An:2020jix,Rashmi:2024ako} for further recent discussion on these states.

\begin{table}
\centering
\caption{Masses (in MeV) of pentaquark states with spin-parity $\left(\frac{5}{2}\right)^-$ for three interactions: central at LO and NLO, and including the hyperfine term.}\label{pentaquarktable}
\begin{tabular}{c|c|c|c|}
\cline{2-4}
& LO     & NLO    & Spin   \\ \hline
\multicolumn{1}{|c|}{{ $(ccc)(c\bar c)$}} & { 7860(200)} & { 7850(200)} & {  7860(200)} \\ \hline
\multicolumn{1}{|c|}{ {  $(ccb)(c\bar c)$}} & 11040(190)                       & 11020(190)                       & 11040(190)                       \\ \hline
\multicolumn{1}{|c|}{ {  $(ccc)(c\bar b)$}} & 11060(190)                       & 11050(190)                       & 11060(190)                       \\ \hline
\multicolumn{1}{|c|}{ {  $(ccb)(b\bar c)$}} & 14240(180)                       & 14210(180)                       & 14240(180)                       \\ \hline
\multicolumn{1}{|c|}{ {  $(bbc)(c\bar c)$}} & 14240(180)                       & 14240(180)                       & 14250(180)                       \\ \hline
\multicolumn{1}{|c|}{ {  $(ccb)(b\bar b)$}} & 17440(170)                       & 17430(170)                       & 17440(170)                       \\ \hline
\multicolumn{1}{|c|}{ {  $(bbc)(c\bar b)$}} & 17450(170)                       & 17450(170)                       & 17450(170)                       \\ \hline
\multicolumn{1}{|c|}{ {  $(bbb)(c\bar b)$}} & 20650(160)                       & 20640(160)                       & 20640(160)                       \\ \hline
\multicolumn{1}{|c|}{ {  $(bbc)(b\bar b)$}} & 20650(160)                       & 20660(160)                       & 20650(160)                       \\ \hline
\multicolumn{1}{|c|}{ {  $(bbb)(b\bar b)$}} & 23830(150)                       & 23830(150)                       & 23830(150)                       \\ \hline
\end{tabular}
\end{table}

\subsubsection{Orders of magnitude for detectability}
Here we give a rule of thumb for the detectability of these multi-heavy-quark states at colliders, particularly  $pp$ at the LHC operating at $\sqrt{s}=13\;\mathrm{TeV}$, since many of the states quoted are beyond the reach of other currently operating particle accelerators.

First, table~\ref{cross sections} gives the rough size of the cross-section $\sigma$ for the production of hadrons involving one or two heavy quarks from several experimentally  known cross sections, and then guesses $\sigma$ for three heavy ones by treating the ratio of the first two due to a multiplicative suppression due to the additional heavy quark.

\begin{table}[h]
\centering
\caption{Order of magnitude of cross sections (in $\mu$barn) for production of basic hadrons containing the stated number of heavy quarks.}
\label{cross sections}
\begin{tabular}{|c|c|l|c|c|}
\cline{1-2} \cline{4-5}
{  $\phantom{cc}c$ \cite{Qin:2020zlg}}             & {  $\sim 10^2$} &  & {  \phantom{bb}$b$ \cite{LHCb:2017vec}} & {  $\sim 10^2$} \\ \cline{1-2} \cline{4-5} 
$\phantom{c}cc$ \cite{Qin:2020zlg}                                   & $\sim 10^{-1}$                     &  & $\phantom{b}bb$ \cite{LHCb:2018yzj}                       & $\sim 5\cdot10^{-2}$               \\ \cline{1-2} \cline{4-5} 
 $ccc$ \cite{GomshiNobary:2006tzy} & $\sim 10^{-4}$                     &  & $bbb$ \cite{GomshiNobary:2006tzy}              & $\sim 10^{-5}$                     \\ \cline{1-2} \cline{4-5} 
\end{tabular}
\end{table}

Particularly, the suppression factor for the  $c$ quark was extracted from the ratio~\cite{Qin:2020zlg}
\begin{equation}
    \frac{\sigma(\Xi_{cc}^{++})}{\sigma(\Lambda_c^+)}B(\Xi_{cc}^{++}\longrightarrow\Lambda_c^+ K^- \pi^+ \pi^+)=(2.2\pm0.6)\cdot 10^{-4}
    \Longrightarrow
    \frac{\sigma(\Xi_{cc}^{++})}{\sigma(\Lambda_c^+)}\sim10^{-3},
\end{equation}
having guessed $B(\Xi_{cc}^{++}\longrightarrow\Lambda_c^+ K^- \pi^+ \pi^+)\approx 1/6$ since there are six contributing decay modes.

A similar procedure has been applied to $bb$, where the production cross sections for the ground- and first excited- states of bottomonium~\cite{LHCb:2018yzj} have been used. 
This is obviously a very rough, order of magnitude estimate, and it is easy to find exceptions. For example, some  $D_c$ mesons, with one $c$-quark only, have cross sections at the mbarn level, \cite{LHCb:2015swx}.

Basically, one can operate under the following empirical rule:
adding a $c$-quark implies a fall-off in the cross section by a factor about $10^3$, 
while adding a $b$-quark costs a larger factor (2000 upon adding the second and 5000 for the third quark). 

Thinking now of pentaquarks such as $(ccc)(c\bar c)$, one would expect cross sections of order $10^{-10}\;\mathrm{\mu b}$, that is, 0.1 fbarn
which might be reachable with the HL-LHC program, and a more unassailable
$10^{-4}\;\mathrm{f b}$ 
for the $(bbb)(b\bar b)$ combination.
The production of other states in this article can be estimated along similar lines. Running full simulations to obtain actual cross-section predictions is far beyond the scope of this work, but the interested reader can find existing works~\cite{Vega-Morales:2017pmm}.

\subsection{Hexaquark/dibaryon color and spin structure. Mass estimate.}
The Hartree-Fock method is specially useful for escalating our method to systems with more heavy quarks. As a final extension, we will now make a brief exploration of the possible hexaquark computations with the same setup.

The first heavy dibaryons that come to mind are the $6c$ or $6b$, those composed of six charm or six bottom quarks, with the color structure of two baryons, i.e. $\frac{1}{6}\varepsilon^{ijk}\varepsilon^{lmn}$. However, this configuration is flavour-symmetric. Then, each constituent baryon must have a symmetric $S=3/2$ spin wavefunction so that Pauli exclusion principle is fulfilled. Since two singlets can not exchange a gluon and remain singlets, we have no $t$-channel contribution, but only the $u$-channel one. The spins between the baryons must be antiparallel to evade Pauli exclusion, so these $u$-channel amplitudes are zero as well. A way around is considering spin-dependent potentials as in Eq.~(\ref{Vspin}).  Since the potential is $1/m^2$ suppressed, these are not very large contributions, so it is plausible that the masses for these hexaquarks will lie very close to the decay threshold into two baryons. We compare the heavy baryon results from \cite{Llanes-Estrada:2011gwu} with the hexaquark masses computed in \cite{G:2024zkc,Alcaraz-Pelegrina:2022fsi} and find this prejudice to be born within the uncertainties in the literature.\\\\
A not so trivial dibaryon system can be studied for heavy `deuterons' composed of charm and bottom quarks, $\mathcal{D}_{bc}$. The constituent baryons will have spin $1/2$, and therefore the $\mathcal{D}_{bc}$ will be in the $J=1$ orbital with $P=(+)$ parity and antisymmetric flavour part $\frac{1}{\sqrt{2}}(\Omega_{ccb}\Omega_{bbc}-\Omega_{bbc}\Omega_{ccb})$, as its Nuclear Physics analog \cite{Junnarkar:2019equ}. The only needed color factor, besides the one for gluon exchange within each constituent baryon which we have previously calculated, is the one of the $u$-channel:
\begin{equation}
    f_{qq} = \frac{1}{36}\varepsilon^{pqr}\varepsilon^{stv} T^a_{si}T^a_{pl}\varepsilon^{ijk}\varepsilon^{lmn}\delta_{jq}\delta_{kr} \delta_{mt} \delta_{nv} = \frac{4}{9}=-\kappa_{c,qq}
\end{equation}
Solving the coupled Hartree-Fock equations for the 6 orbitals yields a mass of $18860(50)$ MeV, which is considerably below the $\Omega_{ccc}\Omega_{bbb}$ and $\Omega_{ccb}\Omega_{bbc}$ thresholds obtained from estimates in the literature \cite{Llanes-Estrada:2011gwu}. Our own computation for these thresholds with the same setup yields 18940(210) MeV and 19020(210) MeV  respectively (table~\ref{baryon1results}). While within the uncertainty band, it is well possible that the heavy deuteron is stable, as the nominal value comes below threshold and we know that our uncertainties are overestimated. Nevertheless, other calculations using the Flux Tube Model \cite{G:2024zkc}, diffusion Monte Carlo \cite{Alcaraz-Pelegrina:2022fsi} and lattice calculations \cite{Junnarkar:2019equ} produce less bound states. This warrants further study.
For more dedicated computations of doubly-heavy dibaryons we refer the reader to other recent work~\cite{Martin-Higueras:2024qaw}.

Once we have established  credible methods for few- and many- quark states which compare reasonably well with the literature, we can proceed to our final contribution, addressing new-physics operators in the $t$-quark sector (which is more sensitive to high-scale effects) from the eventual absence of bound states.

\section{Interactions in the $t$-quark sector} \label{sec:tpotentials}
\subsection{SM potential $V(r)$ }
We first quickly illustrate the procedure to obtain the non-relativistic potential $V(r)$ for different multiquark configurations. Only the ground state will be considered: therefore the spatial wavefunction is symmetric. Furthermore, the color wave function will be a singlet. 

The color interaction has already been described in subsection~\ref{sec:gluon}. 
As for Higgs exchange, the scattering amplitude is
\begin{equation} \label{eq:M_higgs}
    i\mathcal{M} = \bar{u}(q_k) \lb -ig_h \rb u(q_i)\frac{i}{q^2+m_h^2} \bar{v}(\bar{q}_j)\lb -ig_h \rb v(\bar{q}_l)
\end{equation}
where $g_h = m_t/v$ is the Standard Model coupling to fermions, with $m_t$ the mass of the quark top and $v$ the Vacuum Expectation Value (vev) $v = 245$ GeV. In this case, color indices are omitted because the interaction is color-independent. Proceeding as in the gluon case, the short-range potential for a Higgs exchange in a singlet-diquark $V^h_{q\bar{q}}$ is, with $\alpha_h = \frac{g_h^2}{4\pi}$, 
\begin{equation}\label{eq:higgsgeneral}
    V^h_{q\bar{q}}(r) = - \frac{\alpha_h}{r}e^{-m_hr}\ .
\end{equation}
In contrast to the potential of Eq.~(\ref{eq:gluongeneral}), this one is always attractive. This is because the Wick-contraction (-1) factor from $t\leftrightarrow \bar{t}$ is cancelled out by  the (-1) due to spinor exchange ($u\leftrightarrow v$) because $\bar{u}^{(s)}u^{(s)} = u^{(s)\dagger}\gamma^0u^{(s)} \approx \chi^{s\dagger}\chi^s$ while $ \bar{v}^{(s)}v^{(s)} =  v^{(s)\dagger}\gamma^0v^{(s)} = - \chi^{s\dagger}\chi^s$.

The amplitude corresponding to $Z$-boson exchange is in turn
\begin{equation} \label{eq:M_g}
    i\mathcal{M}_{ijkl} = \bar{u}(q_k) \lb -ig_W\gamma^{\mu}\lb v_f - a_f\gamma^5 \rb \rb u(q_i)\\\times \frac{ig_{\mu\nu}}{q^2 + m_Z^2} \bar{v}(\bar{q}_j)\lb -ig_W\gamma^{\nu}\lb v_f - a_f\gamma^5 \rb \rb v(\bar{q}_l) \ ,
\end{equation}
where, for the top quark,
\begin{equation}
\begin{cases}
    g_W = e/2\sin{\theta_W}\cos{\theta_W}\\
    v_f = 1/2 - 4\sin^2{\theta_W}/3\\
    a_f = 1/2\\
\end{cases}    
\end{equation}
Proceeding as in the gluon case, the potential for a Z exchange in quarkonium is
\begin{equation}\label{eq:Zpot}
    V^Z_{s_1s_2s_3s_4}(r) = - \frac{\alpha_W}{r}e^{-m_Zr}\\\times\lb v_f^2\delta_{s_1s_4}\delta_{s_3s_2}  - a_f^2\sigma^i_{s_4s_1}\sigma^i_{s_2s_3}\rb
\end{equation}
Where $\alpha_W = g_W^2/4\pi \sim 0.01$ (with $\sin^2{\theta_W} \sim 0.23$ \cite{CMS:2018ktx}) and $\sigma^i$ are the Pauli matrices. Note that, for a $qq$ ($\bar{q}\bar{q}$) interaction, not only an extra (-1) factor would appear due to Wick contractions, but the $\sigma^i$ indices would be exchanged.

We have not included the photon-mediated interaction because, also contributing a $1/r$ Coulomb piece, its intensity is in the ratio $(1/128)/(0.16)\sim 5\%$ to the strong force. The reasons why we do work with the $Z$ and $h$ bosons that mediate like weak interactions are our intent to reexamine the negative result of Kuchiev, Flambaum and Shuryak concerning the topball, and that those massive exchanges provide a template for the contact interactions beyond the Standard Model to follow.

\subsection{New physics from four $t$-quark  operators in HEFT}\label{subsec:HEFTops}
The complete basis of HEFT-Lagrangian operators has been known~\cite{Brivio:2016fzo,Sun:2022ssa} for a while. 
Our group has previously~\cite{Castillo:2016erh} worked with the coupling between the $t\bar{t}$ system and the  electroweak-symmetry breaking sector ($W$, $Z$, $h$) in the resonance regime, where the Inverse Amplitude Method is employed. Here, we instead concentrate on dimension-6 (the lowest BSM possible) 4-quark contact operators. They can be combined together to give the same interaction for $t$ quarks/antiquarks. We take \textbf{T} = \textbf{U} = 1 in the HEFT Lagrangian~\cite{Brivio:2016fzo} as we have no explicit initial/final state Higgs particles, different operators degenerate into the same effective one when handling only multitop states, so we can read off
\begin{eqnarray}
    \mathcal{L}^{\text{HEFT}}_{4t}\!  =\! \frac{16\pi^2}{\Lambda^2}\! \bigg[ \!\lb r_1\! +\! r_3\! +\! r_4 \rb \!\bar{Q}_LQ_R\bar{Q}_LQ_R \! \ +\  \nonumber
 \nonumber  \lb r_9 + r_{10} + r_{11} \rb  \bar{Q}_L\gamma^{\mu}Q_L\bar{Q}_L\gamma_{\mu}Q_L + \\
     \nonumber  \lb r_{13} + r_{14} + r_{15} \rb \bar{Q}_R\gamma^{\mu}Q_R\bar{Q}_R\gamma_{\mu}Q_R \ +\  
     \nonumber   \lb\! r_{17} + r_{18} + r_{19} + r_{20} \! \rb \bar{Q}_L\gamma^{\mu}Q_L\bar{Q}_R\gamma_{\mu}Q_R  +\\ 
     \nonumber  \lb r_5 + r_7 + r_8 \rb \bar{Q}_L\lambda^AQ_R\bar{Q}_L\lambda^AQ_R \ +\ 
     \nonumber  \lb\! r_{22}\! +\!\! r_{23}\! +\! r_{24}\!\! +\! r_{25}\! \rb\! \bar{Q}_L\gamma^{\mu}\lambda^{\! A}Q_L\bar{Q}_R\gamma_{\mu}\lambda^{\! A}Q_R  + h.c \bigg]\ .
    \\  \label{eq:lagrangianinfull}
\end{eqnarray}
(The customary $C_i$  Wilson coefficients correspond to  $(4\pi)^2 r_i$ in this notation of \cite{Brivio:2016fzo}, and incidentally, at this level the SMEFT and HEFT operators are equivalent.)

To shorten the notation, we rename the coefficient combinations, in the order in which they appear, $\tilde{K}_1:=16\pi^2\lb r_1 + r_3 + r_4  \rb/\Lambda^2$, $\tilde{K}_2=16\pi^2\lb r_9 + r_{10} + r_{11} \rb/\Lambda^2$ up to $\tilde{K}_6$. 

Given the scarce data, we will limit ourselves to operator combinations respecting $C$ and $P$ symmetries.
Considering scalar, vector, colored-scalar and colored-vector interactions, the effective Lagrangian which we reasonably believe can be constrained with searches for top/antitop bound states at this point would have the following form,
\begin{equation}\label{eq:BSMlagrangian}
    \mathcal{L}^{\text{BSM}}_{4t} = K_1 \lb\bar{Q}Q\rb\!\lb\bar{Q}Q\rb + K_2\lb\bar{Q}\gamma_{\mu}Q\rb\!\lb\bar{Q}\gamma^{\mu}Q\rb+ K_5 \lb\bar{Q}\lambda^AQ\rb\!\lb\bar{Q}\lambda_AQ\rb + K_6\lb\bar{Q}\gamma_{\mu}\lambda^AQ\rb\!\lb\bar{Q}\lambda_A\gamma^{\mu}Q\rb \ .
\end{equation}

This reduced Lagrangian with coefficients $K_i$ (of dimension $E^{-2}$) should be matched to that of SMEFT/HEFT in Eq.~(\ref{eq:lagrangianinfull}).  To do it we need to note  that Equation (\ref{eq:lagrangianinfull}), the HEFT Lagrangian, is written in the chiral basis but  Equation (\ref{eq:BSMlagrangian}) is not. 
For example, 
\begin{equation}
K_1 \bar{Q}Q\bar{Q}Q = K_1\lb \bar{Q}_LQ_R\bar{Q}_LQ_R + \bar{Q}_LQ_R\bar{Q}_RQ_L +h.c.\rb
\end{equation}
so that $K_1$ is matched to $\tilde{K}_1$, and is thus constrained by $C_1+C_3+C_4$, but is also matched to another combination which should equal $\tilde{K}_1$ given our more symmetric restricted Lagrangian. 
Likewise, decomposing $K_2\bar{Q}\gamma_\mu Q\bar{Q}\gamma^\mu Q$ by plugging in $Q=Q_L+Q_R$ we see that $K_2$ must give the term with $\tilde{K}_2$, and is thus constrained by $(r_9+r_{10}+r_{11})$ but also that with $\tilde{K}_3$ and twice that with $\tilde{K}_4$, so it is also constrained by $(r_{13}+r_{14}+r_{15})$ and $\frac{1}{2}(r_{17}+r_{18}+r_{19}+r_{20})$.  Then, $K_5$ is directly matched to $\tilde{K}_5$ among others, and is therefore directly constrained by $(r_5+r_7+r_8)$, and finally $K_6$ matches to $\tilde{K}_6$ and is constrained by $(r_{22}+r_{23}+r_{24}+r_{25})$.
Further constraints on the $K_i$ are less straightforward because of the Fierz identities linking various operators, such as~\cite{Brivio:2016fzo} 
\begin{eqnarray}
    \lb\bar{Q}_LQ_R\rb\lb\bar{Q}_RQ_L\rb = - \lb\bar{Q}_L\gamma_{\mu}Q_L\rb\lb\bar{Q}_R\gamma^{\mu}Q_R\rb/2 \\
 \lb \bar{Q}_L\gamma_\mu \lambda^A Q_L\rb^2 = \frac{1}{3}  \lb  \bar{Q}_L\gamma_\mu  Q_L\rb^2 + 
  \lb  \bar{Q}_L\gamma_\mu \sigma^j Q_L\rb^2
\end{eqnarray}
The relations among the $\tilde{K}_i$ coefficients needed to be able to match Eq.~(\ref{eq:lagrangianinfull}) and the $CP$-invariant Eq.~(\ref{eq:BSMlagrangian}) can then be given as 
\begin{equation}
\tilde{K}_1 = -\tilde{K}_4 \ ; \ \ \  \
\tilde{K}_2=\tilde{K}_3 = \tilde{K}_4/4\ ; \  \ \  \
\tilde{K}_5=-\tilde{K}_6 
\end{equation}
and in the last column of table~\ref{tab:operators_sb} we list the constraints on the $C_i=(4\pi)^2 r_i$ coefficients of the SMEFT operators
\begin{equation}
\begin{cases}
\mathcal{O}_{tt}^1 = \bar{Q}_R\gamma^{\mu}Q_R\bar{Q}_R\gamma_{\mu}Q_R \ \ \ \ \ \ \
\mathcal{O}_{QQ}^1 = \bar{Q}_L\gamma^{\mu}Q_L\bar{Q}_L\gamma_{\mu}Q_L \\
\mathcal{O}_{Qt}^1 = \bar{Q}_L\gamma^{\mu}Q_L\bar{Q}_R\gamma_{\mu}Q_R \ \ \ \ \ \ \ \
\mathcal{O}_{Qt}^8 = \bar{Q}_L\gamma^{\mu}\lambda^AQ_L\bar{Q}_R\gamma_{\mu}\lambda^AQ_R \ . \\ 
\end{cases}
\end{equation}

 Comparing with the Lagrangian from Equation (\ref{eq:lagrangianinfull}) it is clear that each SMEFT constraint is applied on $2\tilde{K}_3$, $2\tilde{K}_2$, $2\tilde{K}_4$ and $2\tilde{K}_6$, respectively (the factors 2 are due to Hermitian conjugate terms). Taking into account the relations among the $\tilde{K}_i$, and the most restrictive constraint for each, the $\tilde{K}_i$ are constrained in the 5th column of Table \ref{tab:operators_sb}.\\
Experimental constraints from CMS~\cite{CMS:2019jsc} which apply to these new-physics operators (interpreted in SMEFT) are illustrated in Table \ref{tab:operators_sb}. If more than one constraint on the same coefficient is possible, the most restrictive one has been chosen.
\begin{table*}
    \centering
    \caption{Four-quark operators in the HEFT basis and their Wilson Coefficients. All constraints taken at a scale of $\Lambda = 1$ TeV.
    The allowed ranges in the rightmost column, from which those on the third to last one are derived, correspond to 95\% confidence level and are taken from \cite{CMS:2019jsc} for the corresponding Wilson coefficient $C_i=(4\pi)^2r_i$.}
    \begin{tabular}{|ccccc||cc|} \hline
        Type & Operator & Composition & Coefficient & 
       Allowed range
        & SMEFT& 
        Range  \\
        & & & &  (TeV$^{-2}$) & op.  & (TeV$^{-2}$) 
        \\\hline 
         Scalar & $\mathcal{O_{S}}$ & $\bar{Q}Q\bar{Q}Q$ & $K_1$, -$K_4$ & [-1.8,1.9] & $\mathcal{O}_{tt}^1$ & [-2.2,2.1] \\
        
         Vectorial &$ \mathcal{O_{V}} $ &$\bar{Q}\gamma^{\mu}Q\bar{Q}\gamma_{\mu}Q$ & $K_2$, $K_3$, $K_4/4$& [-0.5,0.4] & $\mathcal{O}^1_{QQ}$ & [-2.2,2.0] \\
         Color$\otimes$ scalar & $\mathcal{O_{SC}}$ & $\bar{Q}\lambda^AQ\bar{Q}\lambda_AQ$ & $K_5$ & [-6.8,8.0] & $\mathcal{O}_{Qt}^1$ & [-3.7,3.5] \\
         Color$\otimes$ vector & $\mathcal{O_{VC}}$ & $(\bar{Q}\gamma^{\mu}\lambda^AQ)^2$ & $K_5$, -$K_6$ & [-6.8,8.0] & $\mathcal{O}_{Qt}^8$ & [-8.0,6.8] \\ \hline
    \end{tabular}
    \label{tab:operators_sb}
\end{table*}
We next circulate over a few of these interactions such as they are needed for nonrelativistic wave equations.

\subsubsection{Scalar contact interaction}
First in table~\ref{tab:operators_sb} are the contact interactions with spin-0 currents and symmetric under $C$ and  $P$.
The scattering amplitude corresponding to this operator is now
\begin{equation}
    i\mathcal{M} = \bar{u}(q_k) u(q_i) \lb -iK_1  \rb\bar{v}(\bar{q}_j)v(\bar{q}_l)
\end{equation}
Performing the Fourier transform in the non-relativistic limit, the potential for such scalar contact interaction in a singlet-$tt$ (or $t\bar{t}$) $V^{Sc}_{q\bar{q}}$ is 
\begin{equation}\label{eq:contactgeneral}
    V^{Sc}_{q\bar{q}} (\mathbf{r}) = - K_1\cdot\delta^{(3)}\lb \mathbf{r} \rb
\end{equation}
As in the Higgs case, the potential always has the same sign, be it attractive or repulsive depending on $K_1$.
\subsubsection{Vector contact interaction}
Next, we consider a vector current interaction. The corresponding amplitude is now
\begin{equation}
    i\mathcal{M} = \bar{u}(q_k) \gamma^{\mu} u(q_i) \lb -iK_2  \rb\bar{v}(\bar{q}_j)\gamma_{\mu}v(\bar{q}_l)
\end{equation}
As in the one-gluon exchange case, only the $\mu=0$ term contributes.\\Performing the Fourier transform in the non-relativistic limit, the potential is 
\begin{equation}\label{eq:contactgeneralvec}
    V^{Vc}_{q\bar{q}} (\vec{r}) = - K_2\cdot\delta^{(3)}\lb \mathbf{r} \rb
\end{equation}
In the $q-q$ (or $\bar{q}-\bar{q}$) case, the potential would have an extra $(-1)$ due to Wick contractions, just as in the gluon case. Therefore, if, for example, $K_2>0$, the interaction would be attractive for quark-antiquark and repulsive for quark-quark or antiquark-antiquark.
\subsubsection{Contact interaction and BSM resonances}\label{contactyukawa}
Contact interactions from Equations (\ref{eq:contactgeneral}) and (\ref{eq:contactgeneralvec}) can be interpreted as the low-energy limit of a massive BSM boson exchange. In the case of a scalar exchange from Equation (\ref{eq:M_higgs}) and in the low-energy limit $M^2>>q^2$ the amplitude is
\begin{equation}
    \mathcal{M} = \lb \bar{u}u \rb \lb \bar{v}v\rb \frac{g^2}{M^2}
\end{equation}
Using the Born approximation, a contact-like potential is obtained
\begin{equation}
    V(\mathbf{r}) = \frac{g^2}{M^2}\delta^{(3)}(\mathbf{r}) \ .
\end{equation}
With this result, we can relate the Wilson coefficients $C_i/\Lambda^2$ to the coupling of the BSM interaction $g$
and the effective field theory scale $\Lambda$ to the mass of the exchanged particle $M$,
\begin{equation}
     C_i \sim g^2\ \ \ \ \ \Lambda \sim M
\end{equation}

\section{States of $N\geq 2$ top quarks or antiquarks in Hartree approximation}
Because the number of interacting pairs grows as $N(N-1)/2$ and that of coordinates as $3N-5$, multiquark state combinations are candidates for mean-field treatments such as the Hartree-Fock method, or the simpler Hartree approximation here employed. The interactions in systems with $N>2$ top quarks will be the sum of two-quark (antiquark) interactions, therefore, the potential between two quarks $i$ and $j$, $V_{ij} (r)$, will be the same as those of Equations (\ref{eq:gluongeneral}), (\ref{eq:higgsgeneral}), (\ref{eq:contactgeneral}) and (\ref{eq:contactgeneralvec}). However, in the gluon-exchange case, the color factor $C_F$ will depend on the color wave function of the system, which depends on $N$ and the $SU(3)_C$ representation. The color wavefunctions and color factors \cite{Park:2013fda}  are summarized in Table \ref{tab:colorfactors}. 

\begin{table*}
    \centering
    \caption{Wavefunctions and color factors for color-dependent interaction. The superscripts and subscripts $CV$ stand for color vector interactions, $CS$ for color scalar interactions, $V$ for vector color-independent interactions and $S$ for scalar color-independent interactions }
    \begin{tabular}{|c|c|c|c|c|c|c|c|c|c|}
    \hline
        State & $SU(3) $ rep. & Color wavefunction & Participants & $F_{ij}^{CV}$ & $F_{ij}^{CS}$ & $\eta_{CV}$ & $\eta_{CS}$ & $\eta_{V}$ & $\eta_{S}$\\\hline
        
        $t\bar{t}$ & 1 & $\frac{1}{\sqrt{3}}\delta_{ij}\ket{q_i\bar{q}_j}$ & $q\bar{q}$ & $-\frac{4}{3}$ & $-\frac{4}{3}$ & $-\frac{4}{3}$ & $-\frac{4}{3}$ & -1 & -1 \\\hline
        
        \multirow{4}{*}{$2t2\bar{t}$} & \multirow{2}{*}{3$\times \bar{3}$ }&\multirow{2}{*}{$\frac{1}{\sqrt{12}}\epsilon^{aij}\epsilon_{akl} \ket{q_i q_j \bar{q}^k \bar{q}^l }$ }& $qq$ & -$\frac{2}{3}$ & $+\frac{2}{3}$ & \multirow{2}{*}{$-\frac{4}{3}$}  & \multirow{2}{*}{$0$} & \multirow{2}{*}{-2}&\multirow{2}{*}{-3} \\\cline{4-6}
         
         & & &$q\bar{q}$ & $-\frac{1}{3}$  & $-\frac{1}{3}$ & &  & &\\\cline{2-10}
          & \multirow{2}{*}{6 $\times \bar{6}$ }& \multirow{2}{*}{$\frac{1}{\sqrt{6}}d^{aij}d_{akl} \ket{q_i q_j \bar{q}^k \bar{q}^l }$ }  & $qq$ &  $+\frac{1}{3}$ & $-\frac{1}{3}$ &\multirow{2}{*}{$-\frac{4}{3}$}   & \multirow{2}{*}{$+\frac{4}{3}$} & \multirow{2}{*}{-2} &\multirow{2}{*}{-3} \\\cline{4-6}
         
         & & & $q\bar{q}$ & -$\frac{5}{6}$ & $-\frac{5}{6}$ & & & &\\\hline
         
         $3t$ & 1 & $\frac{1}{\sqrt{6}}\epsilon^{ijk}\ket{q_iq_jq_k}$ & $qq$ & $-\frac{2}{3}$ & $+\frac{2}{3}$ & $-\frac{4}{3}$ & $+2$ & +2 & -2\\\hline
         
         \multirow{7}{*}{$3t3t$} & \multirow{2}{*}{1 $\times$ 1}&$\frac{1}{\sqrt{36}}\epsilon_{ijk}\epsilon_{lmn}$ & Within singlet & -$\frac{2}{3}$ & $+\frac{2}{3}$ &\multirow{2}{*}{$-\frac{4}{3}$}&\multirow{2}{*}{$+\frac{4}{3}$} & \multirow{2}{*}{+5} & \multirow{2}{*}{-5}  \\\cline{4-6}
         
         & &$ \times\ket{q_i q_j q_kq_lq_mq_n}$ & Singlet-Singlet & 0 & 0 & & & & \\\cline{2-10}
          & \multirow{5}{*}{8 $\times$ 8\footnote{$T^{(a)}\rightarrow \epsilon^{(a)}$ stands for interaction between quark $i$ and quark $j$ or $k$, across the antitriplet, $\epsilon^{(a)}\rightarrow\epsilon^{(a)}$ for interaction between quark $j$ and quark $k$, both in the same antitriplet, etc. } }&\multirow{5}{*}{$\frac{1}{\sqrt{8}} T^a_{ir}\epsilon_{rjk}T^a_{lp}\epsilon_{pmn}$ }& $T^{(a)}\rightarrow\epsilon^{(a)}$ & $\frac{5}{6}$ & $-\frac{5}{6}$ &\multirow{5}{*}{$-\frac{4}{3}$} & \multirow{5}{*}{$+\frac{4}{3}$}  & \multirow{5}{*}{$+5$} & \multirow{5}{*}{$-5$}  \\\cline{4-6}
          & & $\epsilon^{(a)}\rightarrow\epsilon^{(a)}$ & 
          $\times\ket{q_i q_j q_kq_lq_mq_n}$
          & $-\frac{2}{3}$ & $\frac{2}{3}$ & & & &\\\cline{4-6} 
          & & & $T^{(a)}\rightarrow T^{(b)}$ & $-\frac{1}{3}$ &$\frac{1}{3}$&  & & & \\\cline{4-6}
          & & & $T^{(a)}\rightarrow \epsilon^{(b)}$ & -$\frac{7}{12}$ &$\frac{7}{12}$&  &  & &\\\cline{4-6}
          & & & $\epsilon^{(a)}\rightarrow\epsilon^{(b)}$ & $-\frac{5}{6}$ & $\frac{5}{6}$& & & & \\\hline

          \multirow{8}{*}{$6t6\bar{t}$\footnote{The third column must be multiplied by $\ket{q_iq_jq_kq_lq_mq_n\bar{q}_{c}\bar{q}_{t}\bar{q}_{v}\bar{q}_{w}\bar{q}_{y}\bar{q}_{z}}$}} & \multirow{2}{*}{$1^{\times 4}$}& \multirow{2}{*}{ $\frac{1}{36}\epsilon_{ijk}\epsilon_{lmn}\epsilon^{ctv}\epsilon^{wyz}$} & Within singlet & -$\frac{2}{3}$ & $+\frac{2}{3}$ &\multirow{2}{*}{$-\frac{4}{3}$}&\multirow{2}{*}{$+\frac{4}{3}$} & \multirow{2}{*}{+5} & \multirow{2}{*}{-5}  \\\cline{4-6}
         
         & & & Singlet-Singlet & 0 & 0 & & & & \\\cline{2-10}
          & \multirow{6}{*}{$8^{\times 4}$}&\multirow{6}{*}{$\frac{1}{8} T^a_{ir}\epsilon_{rjk}T^a_{lp}\epsilon_{pmn}$ }& $T^{(a)}\rightarrow\epsilon^{(a)}$ & $\frac{5}{6}$ & $-\frac{5}{6}$ &\multirow{6}{*}{$-\frac{4}{3}$} & \multirow{6}{*}{$+\frac{4}{3}$}  & \multirow{6}{*}{$-6$} & \multirow{6}{*}{$-11$}  \\\cline{4-6}
          & &$\times T^b_{cd}\epsilon_{dtv}T^b_{wx}\epsilon_{xyz}$ & $\epsilon^{(a)}\rightarrow\epsilon^{(a)}$ & $-\frac{2}{3}$ & $\frac{2}{3}$ & & & &\\\cline{4-6} 
          & & & $T^{(a)}\rightarrow T^{(b)}$ & $-\frac{1}{3}$ &$\frac{1}{3}$&  & & & \\\cline{4-6}
          & & & $T^{(a)}\rightarrow \epsilon^{(b)}$ & -$\frac{7}{12}$ &$\frac{7}{12}$&  &  & &\\\cline{4-6}
          & & & $\epsilon^{(a)}\rightarrow\epsilon^{(b)}$ & $-\frac{5}{6}$ & $\frac{5}{6}$& & & & \\\cline{4-6}
          & & & $q\bar{q}$ &0  & 0& & & & \\\hline
    \end{tabular}
    \label{tab:colorfactors}
\end{table*}

The Hamiltonian for the N-top quark system will be taken to have two-body potentials $V_{ij}$
whereas the wavefuncion can be expanded, in the Hartree approximation, as the product of one-particle orbitals in Eq.~(\ref{eq:prod_spinorbitals})
With the goal of setting a first constraint from bound-state physics on the BSM coefficients, it is enough to consider this Hartree variational approximation. Its improvement to the Hartree-Fock one from the Slater determinant in Eq.~(\ref{Slaterdet}) is left for future work.

The spatial part is totally symmetric under the exchange of any quark pair and we will take variations respect to the function $\phi(\textbf{r}_N)$. The color wavefunctions will be taken from Table \ref{tab:colorfactors} and the spin wavefunction will be constructed in order to make the whole wavefunction antisymmetric and with total spin $S=0$ for mesons, $S=3/2$ for baryons.

Introducing the wavefunction of Eq.~(\ref{eq:prod_spinorbitals}) in the variational  Eq.~(\ref{variational}) results in
\begin{equation}\label{eq:var}
    0 = \bra{\delta\phi(\textbf{r})}\left[ -\frac{N}{2m_t} \nabla^2 \ket{\phi(\textbf{r})} + N\eta W(\textbf{r})\ket{\phi(\textbf{r})}   - N\epsilon_i\ket{\phi(\textbf{r})}\right] 
\end{equation}
where the Hartree potential $W(\textbf{r})$ is given by
\begin{equation}
    W(\textbf{r}) = \int d^3\textbf{r}' \abs{\phi(\textbf{r}')}^2 V(\abs{\textbf{r}-\textbf{r}'})
\end{equation}
and $\eta$ is a numerical factor that depends on the interaction being considered viz., for the QCD interaction, also on the color overlap.
Should spin interactions be included, $\eta$ would also depend on the spin wavefunction.
In the case of color interactions, $\eta$ is given by the sum of the color factors of any equally contributing diagrams; its values are listed in table~\ref{tab:colorfactors}.

\subsection{2$t$2$\bar{t}$ tetraquark configurations}
Let us give one detailed example of the treatment in this section: that of  a one-gluon-exchange interaction within the tetraquark in the triplet $\times$ antitriplet representation of $SU(3)_C$. 
This is less far-fetched than it would sound as four-$t$ spectra are now being recorded at the LHC~\cite{VandenBossche:2024trx,ATLAS:2023ajo,CMS:2023zdh}.

The potential operator is, following the notation of the first row of Table \ref{tab:colorfactors}, $V = V_{ij} + V_{ik} + V_{il} + V_{jk} + V_{jl} +V_{kl}$. Therefore, the matrix element of $V$ is
\begin{equation}
    \bra{\Psi} V \ket{\Psi} = \lb \frac{-2}{3}\!\times\! 2 -\frac{1}{3}\!\times\! 4 \rb  \bra{\phi(\textbf{r})\phi({\textbf{r}'})} \frac{\alpha_s}{\textbf{r}-\textbf{r}'}\ket{\phi(\textbf{r})\phi({\textbf{r}'})} = \frac{-8}{3}\bra{\phi(\textbf{r})\phi({\textbf{r}'})} \frac{\alpha_s}{\textbf{r}-\textbf{r}'}\ket{\phi(\textbf{r})\phi({\textbf{r}'})}
\end{equation}
Applying Leibnitz's rule and using that $\textbf{r}$, $\textbf{r}'$ are dummy variables, $\delta \bra{\Psi} V \ket{\Psi} = 0$ yields
\begin{equation}
    0 =\bra{\delta{\phi}}\times\left[ - \frac{16}{3}\bra{\phi(\textbf{r}')} \frac{\alpha_s}{\textbf{r}-\textbf{r}'}\ket{\phi(\textbf{r}')}\right]\ket{\phi(\textbf{r})}
\end{equation}
(In this case, $ N\eta=-16/3 $, see Equation (\ref{eq:var})). Putting all together, we arrive to the Hartree-Fock equation for tetraquark in the triplet $\times$ antitriplet color representation for one-gluon exchange: 
\begin{equation}\label{eq:tetraquark_hf}
    -\frac{1}{2m_t}\nabla^2\phi(\textbf{r}) - \frac{4\alpha_s}{3}\int d\textbf{r}' \abs{\phi(\textbf{r}')}^2 \frac{1}{\abs{\textbf{r}-\textbf{r}'}}\phi(\textbf{r}) = \epsilon\phi(\textbf{r})
\end{equation}
Once again, the binding energy will be given by Koopman's Theorem \cite{franciscoblanco} for the $N$ particles in the same spatial orbital $\phi$, in the form ($h$ being the one-body part of the Hamiltonian)
\begin{equation}\label{eq:koopman}
    E = \frac{N}{2}\left ( \epsilon + \int d^3\textbf{r}' \phi(\textbf{r}')^{*} h \phi(\textbf{r}')  \right )
\end{equation}
However, this incorrectly counts the center of mass motion which should not enter a computation of the mass. This can again be partially amended by correcting the kinetic energy of each particle \cite{DeGregorio:2021dsr} by $-p_i^2/(2mN)$. Therefore, the general self-consistent equation for an $N$ top-quark system with an interaction $V(r)$ will be given by
\begin{equation}\label{eq:hf}
    -\frac{1}{2m_t}\left( 1 - \frac{1}{N} \right)\nabla^2\phi(\textbf{r}) +\eta\int d^3\textbf{r}' \abs{\phi(\textbf{r}')}^2 V(\abs{\textbf{r}-\textbf{r}'})\phi(\textbf{r}) = \epsilon\phi(\textbf{r})
\end{equation}
The factor $\eta$ is calculated by applying the procedure followed to arrive to Equation (\ref{eq:tetraquark_hf}). For a state of $N$ $t$ quarks, the general formula for $\eta$ is
\begin{equation}\label{eq:eta}
   \eta = \frac{2}{N}\lb \sum_{i=1}^N\sum_{j>i} F_{ij} \rb\ ,
\end{equation}
where $F_{ij}$ is the numerical factor in the interaction between particle $i$ and particle $j$. For scalar color-independent interactions, $F_{ij} = -1\  \forall i,j$; for vector color-independent interactions $F_{ij} = +1 $ if $i,j$ are both quarks (or antiquarks) or $F_{ij} = -1$ for mixed baryon number. For vector and scalar color-dependent interactions the $F_{ij}$ are given in Table \ref{tab:colorfactors}.

\section{Numerical binding energies for multitop states} \label{sec:numerictop}
In this section we will present the Hartree-Fock predictions for the energies from Equations (\ref{eq:koopman}) and (\ref{eq:hf}). The top quark parameters that have been used are summarized in Table \ref{tab:top}. Note that the strong coupling has been  self-consistenly calculated to one loop at a soft scale, that is, the inverse of the bound state radius \cite{Llanes-Estrada:2011gwu} $r^{-1} \sim m\alpha_s$ which corresponds to typical momenta.
\begin{table}
    \centering    
    \caption{Numerical parameters for the $t$ sector.}
    \begin{tabular}{|ccc|}\hline 
    Magnitude & Value & Ref.\\ \hline
         $m_t$& $171 \pm 2$ GeV & \cite{topmass} \\
         $\Gamma_t$& $1.36\pm0.16$ GeV & \cite{topmass} \\
         $\alpha_{s}(m_t\alpha_s)$& $0.16$ & -
         \\
         $g_{h}$& $1.16\pm0.35$ & \cite{yukawatop}\\ \hline 
    \end{tabular}
    \label{tab:top}
\end{table}
Other states listed in what follows are treated in an entirely analogous manner.

\subsection{Higgs exchange}
\begin{figure}
    \centering
    \includegraphics[width=0.6\textwidth]{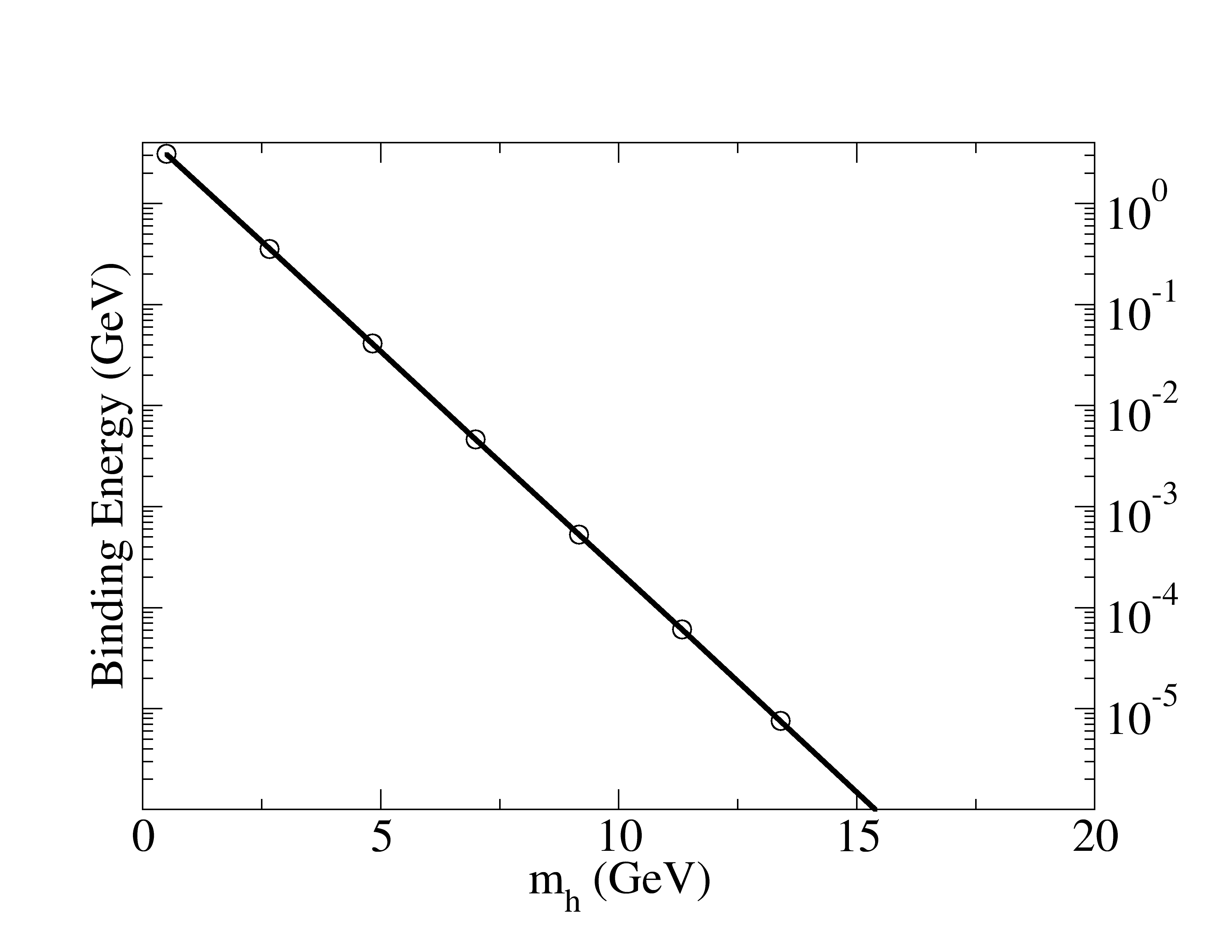}
    \caption{Binding Energy of the 12-particle T-ball in Hartree approximation versus the Higgs boson's mass. The object is less bound by this interaction than by that of Chromodynamics alone, and for higher, physical Higgs masses probably not even at all. For a vector-like exchange the supposed binding would be even less favorable, as the interaction is not universally attractive but its sign depends on whether $tt$ or $t\bar{t}$ interact.} 
    \label{fig:mhBE}
\end{figure}

Let us start by the simplest potential, scalar exchange. We will let the mass of the exchanged boson vary and proportionally fix its coupling to the $t$ as in the SM.  We then compare the dependence of the bound-state energy on the Higgs boson mass for the $N=12$ T-ball with the published computation of Kuchiev, Flambaum and Shuryak~\cite{Kuchiev:2008fd}. Our result is illustrated in Figure \ref{fig:mhBE} and is in qualitative agreement with that prior calculation.

Just as in Ref. \cite{Kuchiev:2008fd}, we see that a bound state for the T-ball would disappear (in our calculation, exponentially) for Higgs masses much smaller than the physical one. In light of these results, and by looking at the $N$ dependence of Equations (\ref{eq:hf}) and (\ref{eq:koopman}), we conclude that the binding energy would be even smaller in states with $N<12$ top quarks. Hence, we confirm that a multitop state bound by the Standard Model 125-GeV Higgs exchange alone should not exist (the QCD interaction barely binds this and the various other multi-$t$ systems as we show shortly in subsec.~(\ref{gluon_ex}).

A difference with Kuchiev, Flambaum and Shuryak is that we do not reach scalar-boson masses as high as 30 GeV with this level of precision (note in the figure that we follow the binding energy over six orders of magnitude on an exponentially falling slope, which already required some careful study of the grid employed and the continuum/infinite volume limits) but we concur with those authors, even in our basic Hartree approximation, on the absence of the Higgs-bound Topball of Froggatt and Nielsen in the SM. We will later in subsection~\ref{subsec:contactintops} revisit this with BSM interactions.

\subsection{Bound on the Z-boson exchange energy}

We now return to the potential for Z-boson exchange calculated in Equation (\ref{eq:Zpot}). As can be seen, it depends on the spin indices, and thus,  on the spin wavefunction of the system. However, as the potential has the same form as the Higgs-exchange one and the Z-boson mass is known ($m_Z \sim 90$ GeV), it will not be necessary to to distinguish these wavefunctions. This is because, to conclude that no deeply  bound states are formed, an upper bound for the spin factor $F_S$ suffices.\\\\On the one hand, we can constrain the Pauli-matrix product contraction
\begin{equation}
    \sigma^{i}_{s_4s_1}\sigma^i_{s_2s_3} \leq 2 \ .
\end{equation}
The inequality is saturated for $s_4 = s_3 = -1$, $s_1=s_2 = 1$. In consequence, for any quark pair, $\abs{F_S} \leq 2a_f^2 = 0.5$. This gives an upper bound for $\eta$. Setting $F_{ij} = 2a_f^2 \ \forall i,j$ in Equation (\ref{eq:eta}),
\begin{equation}
    \eta \leq \frac{2}{N}2a_f^2N(N-1) \propto N-1
\ .
\end{equation}
Even for $N=12$, the total energy resulting from the Hartree approximation is nonnegative, so we assess that there is no bound state for this nor any smaller $N$.

\subsection{Gluon  exchange}\label{gluon_ex}
In Section \ref{sec:gluon} the potential at LO was calculated and was seen to be a Coulomb-like potential. For the two-body system, the solution for the Schrödinger equation is well known and leads to a bound ``toponium'' system with binding energy
\begin{equation}\label{eq:ogexpect}
    |E_0| = \frac{1}{2}\mu \alpha_{\text{eff}}^2 \approx 1.9 \ \text{GeV} 
\end{equation}
(for $t\bar{t}$, $\alpha_\text{eff} = 4\alpha_s/3$ and $\mu = m_t/2 \approx 85.5$ GeV). On the other hand, because the binding energy is not  sizeable enough to significantly contract the phase space, we expect the width of toponium to be $\Gamma_{t\bar{t}} \approx 2\Gamma_t \approx 2.7$ GeV. 

If we adopt Rayleigh's criterion for detectability \cite{rayleigh}, which demands the binding energy to exceed the width of the state, $\abs{\Delta E}>\Gamma$, we see that it is not satisfied here as $1.9<2.7$ by a significant amount. This criterion, borrowed from optics, is also a reasonable starting point in particle physics, given that the broadest resonance discussed in the basic hadron spectrum, the $\sigma$ or $f_0(500)$ light scalar meson, has a mass of about 450 MeV and a width of 280 MeV, and its very existence was doubted for two decades. So we adopt Rayleigh's criterion as benchmark and will consider any particle failing it as nondetectable (but a recent analysis~\cite{Aguilar-Saavedra:2024mnm} has shown that the toponium signal in the $bbWW$ channel corresponds to 1-4\% in the energy region of interest, depending on the binning, so it might be possible to pick it up at the LHC, just not as a conventional Breit-Wigner resonance).

Therefore, although direct reconstruction of toponium seems difficult, by adding more $t/\bar{t}$ quarks, we could naively expect the binding energy to grow faster ($\sim N(N-1)$) than the width ($\sim N$), so an explicit check is convenient.

The Hartree-approximation results for  those states listed in table~\ref{tab:colorfactors} are spelled out in table~\ref{tab:OGE}, with the bounding provided by one-gluon exchange. The multitop mass is calculated from the binding energy (BE) as $M = Nm_t -|BE|$, its width as $\Gamma = N\Gamma_t$ and the error for the binding energy is taken to be at least 50\% of the correction due to the center of mass kinetic energy (of Eq.~(\ref{eq:hf}) above), falling-off with $N$. (This is because for larger systems, each particle recoils against a very large composite effectively at rest, reducing the error.)
\begin{table}
    \centering    
    \caption{Hartree binding energies for multi-$t$ states with LO and NLO QCD potentials (all entries in GeV). We see that \emph{a)} none is bound enough to satisfy Rayleigh's $|BE|>\Gamma$ criterion and \emph{b)} the binding grows rather slowly, not at all with $N^2$. This is because $\eta_{CV}=-4/3$ independently of $N$ (table~\ref{tab:colorfactors}). One can think of a quark, a color triplet, interacting, globally, with an antitriplet formed by all other particles in the hadron, as the HF approach is meant to yield one-body equations.
    }
    \begin{tabular}{|c|ccc|}\hline
    State & LO BE  & LO+NLO BE  & Width \\\hline
         $t\bar{t}$ & -1.7$\pm$0.4 & -2.4$\pm$0.6 &2.7$\pm$0.3\\
         $3t$ & -1.9$\pm$0.3& -2.6$\pm$0.4  &4.1$\pm$0.5 \\
         $2t + 2\bar{t}$ & -2.3$\pm$0.3 & -3.1$\pm$0.4 & 5.4$\pm$0.6\\
         $6t$ & -3.0$\pm$0.2 & -4.2$\pm$0.4 & 8.2$\pm$1.0  \\
         $6t + 6\bar{t}$ & -5.5$\pm$0.2 & -7.6$\pm$0.3 & 16.3$\pm$1.9 \\ \hline
    \end{tabular}
    \label{tab:OGE}
\end{table}

As can be seen in the first entry of Table \ref{tab:OGE}, the Hartree calculation for toponium at LO yields a satisfactory result when compared with the exact solution of the Schr\"odinger equation for the two-body problem (see Eq.~(\ref{eq:ogexpect})), also suggesting that the assigned uncertainty might be too conservative. Comparing the middle and right columns we conclude that no multi-top state bound by one-gluon exchange is expected to be narrow enough since their binding energies do not overcome the large decay rate. As usual in a mean-field treatment of QCD, $BE \sim N$ and not $N^2$ as explained in the caption of table~\ref{tab:OGE}.

\subsection{Contact interaction only} \label{subsec:contactintops}
For contact potentials, the two-body problem is analytically solvable within an adequate renormalization scheme, which allows a rough calibration of the Hartree-Fock method. The analytical solution for the Schrödinger equation with a potential $ V(\textbf{r}) = -g\delta^{(3)}(\textbf{r})$ with coupling constant $g = C/\Lambda^2$, in terms of  the Wilson coefficient $C$ from Equation (\ref{eq:contactgeneral}) or (\ref{eq:contactgeneralvec}),
\begin{equation}
    -\frac{1}{2\mu}\nabla^2\phi(\textbf{r}) - g\delta^{(3)}(\textbf{r})\phi(\textbf{r}) = \varepsilon\phi(\textbf{r})
\end{equation}
is straightforward. Multiplying by $\mu$, defining $\beta \equiv \mu g$, $E \equiv \mu\varepsilon$ and writing $\phi(\textbf{r})$ as the Fourier transform of $\hat{\phi}(\textbf{k})$, one has
\begin{equation}
    \hat{\phi}(\textbf{k}) = \frac{\beta\phi(0)}{\frac{k^2}{2} - E} \ .
\end{equation}
Performing an integration over $\textbf{k}$ and writing $E = -\abs{E}$ in order to solve for bound states,
\begin{equation}
    \frac{(2\pi)^3}{\beta} = 8\pi\int_{0}^{\infty} dk \left[ 1-  \frac{\abs{E}}{\frac{k^2}{2}+\abs{E}} \right]
\end{equation}
This radial integral is divergent, thus requiring regularization. Imposing a cutoff\footnote{This cutoff will be the same as the scale parameter for SMEFT, that is, $\Lambda = 1$ TeV, since it naturally represents a momentum scale at which new physics might manifest.} $\Lambda$ for the integral, the final result is a transcendental equation for $\varepsilon= \abs{E}/\mu$
\begin{equation}\label{eq:trascendental}
    \sqrt{2\varepsilon\mu}\arctan{\lb\frac{\Lambda}{\sqrt{2\varepsilon\mu}}\rb} = \Lambda\lb 1-\frac{\Lambda}{\mu}\frac{\pi^2}{C} \rb  \ .
\end{equation}
A bisection method has been used to solve this transcendental equation for $m_{\text{top}} = 171$ GeV.
For the smallest value of $C$ that yields a solution, the binding energy is
\begin{equation}
    \varepsilon = -3.66 \ \text{GeV}
\end{equation}
but this value of $C=120$TeV$^{-2}$ is far from the experimental constraints spelled out in Table \ref{tab:operators_sb}; hence, toponium would not meet Rayleigh's criterion in full via BSM interactions alone with current constraints. We will have to numerically examine whether it does when combining these with the SM ones due to one-gluon exchange. 

Beyond the 2-body problem, states with a particle number $N>2$ call for a Hartree-Fock-style calculation. The Hartree equation for the scalar color-independent interaction is (noting that $C\propto 16\pi^2 r_i$)
\begin{equation}\label{eq:hf_delta}
\epsilon\phi(\textbf{r})=
-\frac{1}{2m_t}\left( 1 - \frac{1}{N} \right)\nabla^2\phi(\textbf{r}) -(N-1)\frac{C}{\Lambda^2}\int d^3\textbf{r}' \abs{\phi(\textbf{r}')}^2 \delta^{(3)}(\textbf{r} - \textbf{r}')\phi(\textbf{r}) \ .
\end{equation}
Separating the orbital angular and radial parts,  $\phi(\textbf{r}) = R(r)Y_0^0(\theta,\varphi)$,
\begin{equation}
    -\frac{1}{2m_tr}\left( 1 - \frac{1}{N} \right)\frac{d^2}{dr^2}\left[ rR(r) \right] - \frac{(N-1)C}{4\pi\Lambda^2}\abs{R(r)}^2 R(r) = \epsilon R(r)\ .
\end{equation}

That nonlinear Schrödinger equation exhibits numerical instabilities. To avoid them, and
connecting with the discussion in Section \ref{contactyukawa}, the contact interaction is numerically implemented as the large-$m_h$ limit of a family of Yukawa interactions (Equation (\ref{eq:higgsgeneral})) with strength $\alpha_{\rm contact} = C/(4\pi) $ and mass $M = \Lambda$. We will use this approach in the next subsection~\ref{subsec:SMandContact} in which we combine it with those others from the Standard Model.

\subsection{Standard Model+Contact interactions}\label{subsec:SMandContact}
In this section, all described interactions will be simultaneously included in a Hartree-Fock calculation. The HF equation will be the same as Equation (\ref{eq:hf}), and with the contact interaction represented as a Yukawa potential at a high scale, the potential  $\eta V_{\text{Total}}(\abs{\textbf{r}'-\textbf{r}}) $ there is now given by
\begin{eqnarray}
    \eta V_{\text{Total}}(\abs{\textbf{r}'-\textbf{r}}) &=& \frac{\alpha_{\text{contact}}}{\abs{\textbf{r}'-\textbf{r}}}e^{-\Lambda\abs{\textbf{r}'-\textbf{r}}} - (N-1)\frac{\alpha_{h}}{\abs{\textbf{r}'-\textbf{r}}}e^{-m_h\abs{\textbf{r}'-\textbf{r}}} 
    \nonumber \nonumber \\ & &
    - (N-1)\frac{\alpha_{W}}{\abs{\textbf{r}'-\textbf{r}}}e^{-m_Z\abs{\textbf{r}'-\textbf{r}}} -\frac{4}{3}\frac{\alpha_s}{\abs{\textbf{r}'-\textbf{r}}} +V_{\rm NLO}(\textbf{r}'-\textbf{r})  \label{Vtotal}
\end{eqnarray}
Here, $\alpha_{\text{contact}}$ is given by
$ \alpha_{\text{contact}} = \frac{1}{4\pi} \sum_i \eta_i\frac{K_i}{\Lambda^2} $, with $i =$ \{S, V, CS, CV\}; the Wilson coefficients $K_i$ are constrained in Table \ref{tab:operators_sb}, so we have a feeling for the typical maximum value that this variable can still take, and the pre-factors $\eta_i$ are given in Table \ref{tab:colorfactors}. We can then explore the dependence of the two-body binding energy on $\alpha_{\text{contact}}$.
\begin{figure}[!]
    \centering
    \includegraphics[width=0.5\textwidth]{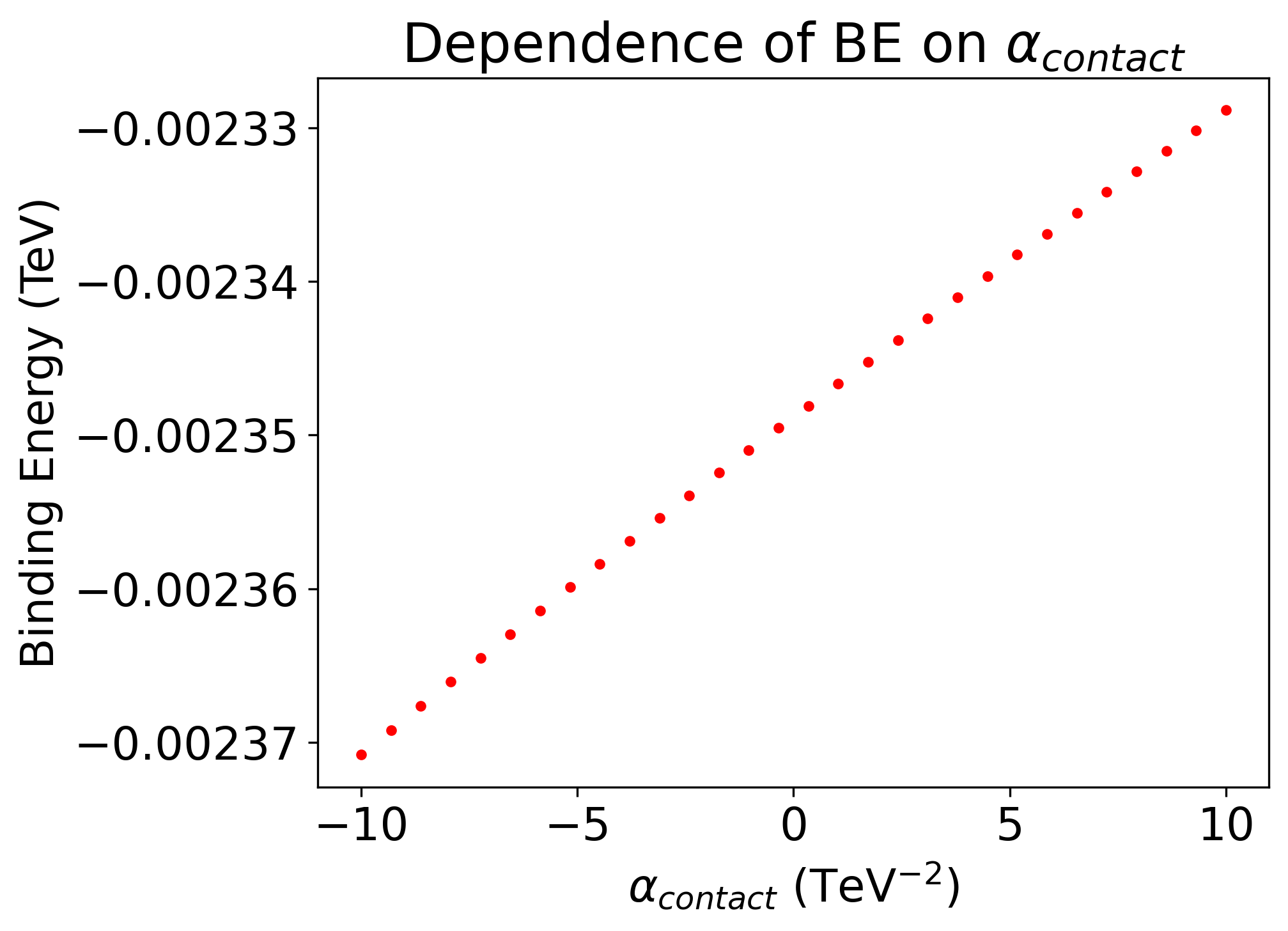}
    \caption{Dependence of the  $N=2$ binding energy on the BSM $\alpha_{\text{contact}}$ with $\Lambda=1$ TeV summed to the SM interactions, as in Eq.~(\ref{Vtotal}).}  
    \label{fig:BEalphaN2}
\end{figure}
In the case of $N=2$, $\alpha_{\text{contact}}$ is given by
\begin{equation}
    \alpha_{\text{contact}} = \frac{1}{4\pi\lb 1 \text{ TeV}\rb^2} \{-C_S  -C_V -\frac{4}{3} C_{CS} -\frac{4}{3}C_{CV} \}
    \label{alphacontactN2}
\end{equation}
Thus, toponium constraints on $\alpha_{\text{contact}}$ only affect this linear combination of Wilson coefficients, and leave orthogonal directions in the ${C_i}$ space unconstrained. 

\subsubsection{Constraints on $\alpha_{\text{contact}}$}
The way to globally constrain all the coefficients from Table \ref{tab:operators_sb} is by looking at the presumed decay spectra of multi-top states: if no peak meeting Rayleigh's criterion is observed at the mass predicted for a fixed $\alpha_{\text{contact}}$, then that value of $\alpha_{\text{contact}}$ is discarded when sufficient numbers of N$t$ events are accumulated. Nowadays, toponium characterisation is still a difficult endeavour \cite{tophunter}, thus, constraints on $\alpha_{\text{contact}}$ are difficult to obtain even for $N=2$. 

Instead, would-be constraints which would follow from eventual undetectability can be proposed, while waiting for data, based on Rayleigh's criteria: the values of $\alpha^{(N)}_{\text{contact}}$ that would yield a binding energy larger than the system's width. These achievable constraints are spelled out in Table \ref{tab:cotas}.
\begin{table}
    \centering
    \caption{Would-be values of $\alpha_{\text{contact}}$ (with a precision of 5 TeV$^{-2}$) needed for detectability, based on Rayleigh's criteria in the absence of bound states when the relevant $N$-particle spectrum becomes accessible. In the third column, the coupling is given by $4\pi\lb1 \text{TeV}\rb^2\alpha_{\text{contact}} = aC_S + bC_V + cC_{CS} + dC_{CV}$.
    The absence in the spectrum of the state corresponding to an inequality means that this should be inverted and turned into a constraint; e.g. the absence ot $t\bar{t}$ toponium would suggest $\alpha^{(2)}_{\rm contact}\geq -1.3$.
    The last column is the expected bound on $C_S$ if it were the only nonvanishing source of new physics (that is, $\alpha_{\rm contact}$ divided by $(N-1)$).
    }
    \begin{tabular}{|ccc|c|} \hline
         N& $(a,b,c,d)$ & Upper bound (TeV$^{-2}$) & Limiting $C_S$\\\hline
         2&  $(-1,-1,-4/3,-4/3)$ & $\alpha^{(2)}_{\text{contact}} \leq -57 $  & -57 \\
         3&  $(-2,+2,+4/3,-4/3)$ & $\alpha^{(3)}_{\text{contact}} \leq -102 $ & -51\\
         4 &  $(-3,-2,+4/3,-4/3)$ & $\alpha^{(4)}_{\text{contact}}\leq -120 $ & -40\\
         5 &  $(-3,-2,+4/3,-4/3)$ & $\alpha^{(4)}_{\text{contact}}\leq -135 $ & -33 \\
         6&  $(-5,+5,+4/3,-4/3)$ & $\alpha^{(6)}_{\text{contact}} \leq -141$ & -28 \\
         12&  $(-11,-6,+4/3,-4/3)$ & $\alpha^{(12)}_{\text{contact}} \leq -139 $&  -13 \\
        \hline
    \end{tabular}
    \label{tab:cotas}
\end{table}
As can be seen, each combination of Wilson coefficients becomes differently bounded. In order to bracket the Wilson coefficients $C_i$ themselves, it would be necessary to have several N-top spectra so the inequalities can be disentangled.

For example, reducing ourselves to the two-by-two system composed by the two uncoloured BSM coefficients $(C_S,C_V)$  and two eventual negative results for the two-body toponium and three-body all-top baryon, we would have
\begin{equation}
\label{disentangle1}
\begin{pmatrix}
-1 & -1 \\ -2 & 2
\end{pmatrix} \begin{pmatrix}
C_S \\ C_V
\end{pmatrix} \leq \begin{pmatrix}
B_2 \\  B_3
\end{pmatrix} = \begin{pmatrix}
-57 \\ -102
\end{pmatrix}
\end{equation}
(see table~\ref{tab:cotas} for the numerical coefficients) from which simple inversion~\footnote{Larger systems can be automatically solved with the help of Mathematica's {Reduce[]} command; each linear inequality eliminates half of the space above/below a certain plane, and the region not excluded is a, generally open, Newton polygon characterized by vertices and edges at the intersections of the planes.} returns, for detectability of both the two- and the three-body resonances,
\begin{eqnarray}
C_S &\geq& 54 \\
C_V &\in& [57-C_S,-51+C_S]
\end{eqnarray}
for detectability in Rayleigh's sense; in the absence ot $t\bar{t}$ with $|BE|>\Gamma$, we would then have $C_S\leq 54$.
(If instead of the two- and three-body problem we employ the dibaryon six-body and the dodecatoplet 12-body systems, we can obtain a somewhat better $C_S\leq 18$.)

A sanity check on these number sizes can be performed by using first order perturbation theory to compute the expectation value of the potential of Eq.~(\ref{eq:contactgeneral}) 
with the hydrogenoid wavefunction of Eq.~(\ref{ansatze+e-}) but with the Bohr radius being $a=(\alpha_s m_t/2)^{-1}$. The value of $K_1$ which, if there is no other operator sourcing new physics, is equivalent to $C_S$ for the two-body system as $-a=(N-1)=1$,  can then be obtained from requiring that Rayleigh's criterion be met with $K_1$ providing the extra binding. This yields
\begin{equation}
\Gamma_{t\bar{t}}-|BE_{t\bar{t}}| = - \frac{K_1}{\pi a^3} 
\end{equation}
and, with the increased binding needed taken as 0.3 GeV, $\alpha_s=0.16$ and $m_t\sim 171 $ GeV, leads to $K_1=370$TeV$^{-2}$.
This is a factor 6 larger than the value of 57 shown in table~\ref{tab:cotas} (we should not expect perfect agreement between the iterated Hartree-Fock 
approximation and the perturbative expression with the exact two-body wavefunction), but it shows that one would need Wilson coefficients significantly larger than 1.

Since extant constraints, such as for example $|C_S|<1.9$, see table~\ref{tab:operators_sb}, are already tighter by an order of magnitude, the search for multi-top bound states will not yield the most accurate bounds; but we hope that the reader will appreciate how different the systematic errors are here and that definitely these will be independent bindings on the parameters.

\subsection{Resonances of multi-top systems}

\subsubsection{The T-Ball}
As mentioned above in Section \ref{intro}, these new-physics interactions could yield a bound state of $6t+6\bar{t}$ if they were sufficiently large. In order to predict the mass of this hypothetical bound state, a Hartree calculation has been carried out with the minimum value of $\alpha_{\text{contact}}$ allowed by the already existing Wilson coefficients' constraints from Table \ref{tab:operators_sb}. The minimum allowed coupling is $\alpha_{\text{contact}} = -3.15 $ TeV$^{-2}$ for $\Lambda = 1$ TeV and
the result is a binding energy $BE = -9.15$ GeV. As can be seen in Table \ref{tab:OGE}, the T-ball width is $\Gamma_{\text{T-ball}} = 16.3$ GeV, thus, even though the bound state would exist, it would be so broad that, according to Rayleigh's criterion, it would not be directly detectable.
Its effect on off-shell observables might help set constraints on the Wilson coefficients.

\subsubsection{Dependence of the binding energy on $N$}
Finally, we gather the various  structures that could emerge from BSM interactions with couplings near the current experimental limits and that we have discussed. 

\begin{figure}
    \centering
    \includegraphics[width=0.48\linewidth]{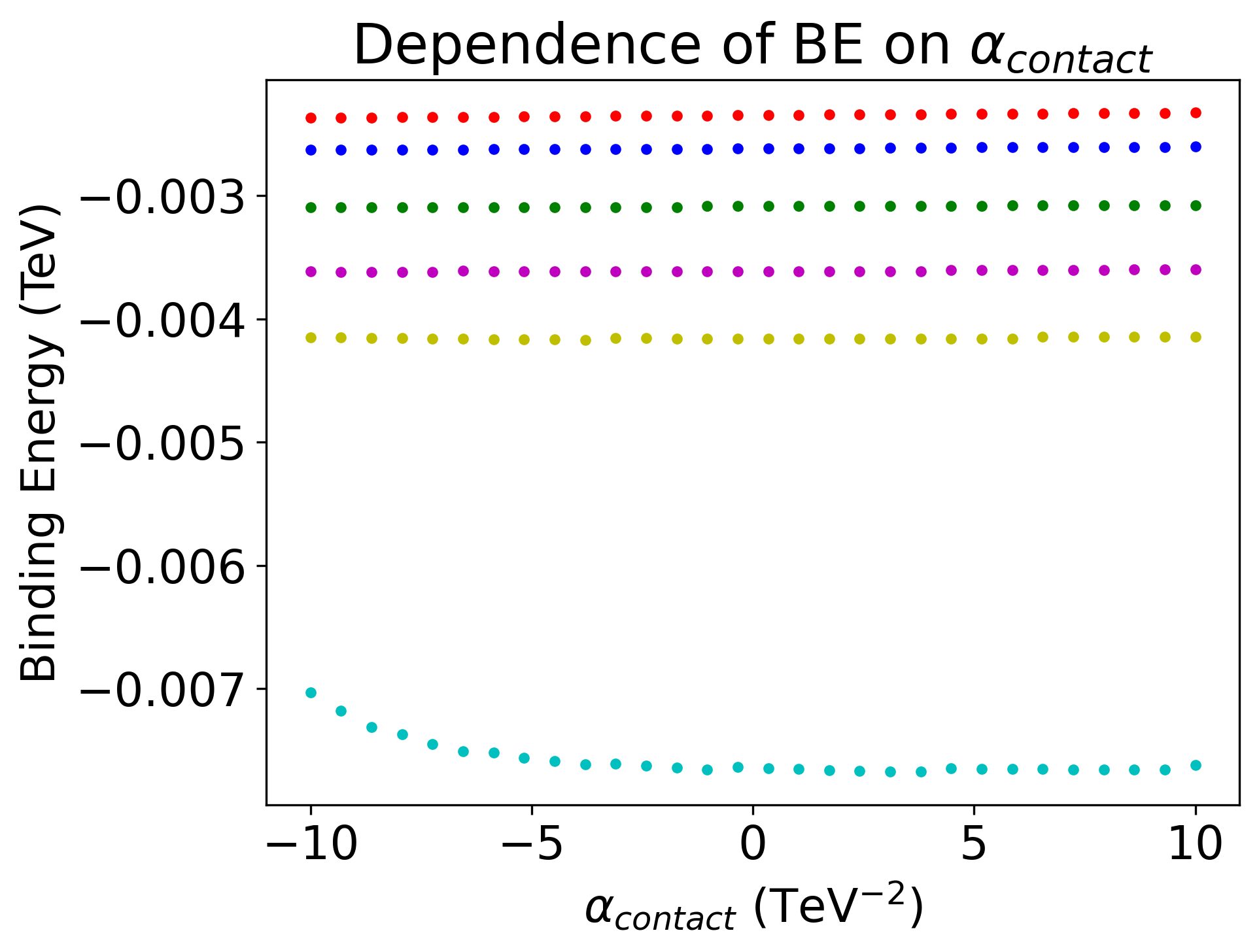}
    \caption{Binding energy (OY axis) computed with SM interactions and contact interaction (coupling on the OX axis) for $N=2,3,4,5,6,12$ (from top to bottom). As can be seen, the dependence with $N$ is not quadratic, and the dependence with $\alpha_{\text{contact}}$ not very steep for these values.
    }
    \label{fig:Ntopbinding}
\end{figure}
In figure~\ref{fig:Ntopbinding} we plot the binding energy as a function of the contact-interaction strength and the number of particles. As can be seen, $BE$ is about linear in $N$, not quadratic (revealing a  hadron picture in which a quark interacts with the rest of the object as a whole, natural in Hartree and Hartree-Fock computations), and the dependence on $\alpha_{\rm contact}$ not very steep for these moderate values (at larger ones it does accelerate, but they are well beyond current bounds on the Wilson coefficients of HEFT, so we believe it to be idle to plot them).  For orientation, with the extreme values for the Wilson coefficients in table~\ref{tab:operators_sb}, we would expect $|\alpha_{\rm  contact}|<2$ or thereabout (remember the $1/(4\pi)$ factor in Eq.~(\ref{alphacontactN2})), well within the graph, that runs between -10 and 10.

\begin{figure}
    \centering
    \includegraphics[width = 0.5\textwidth]{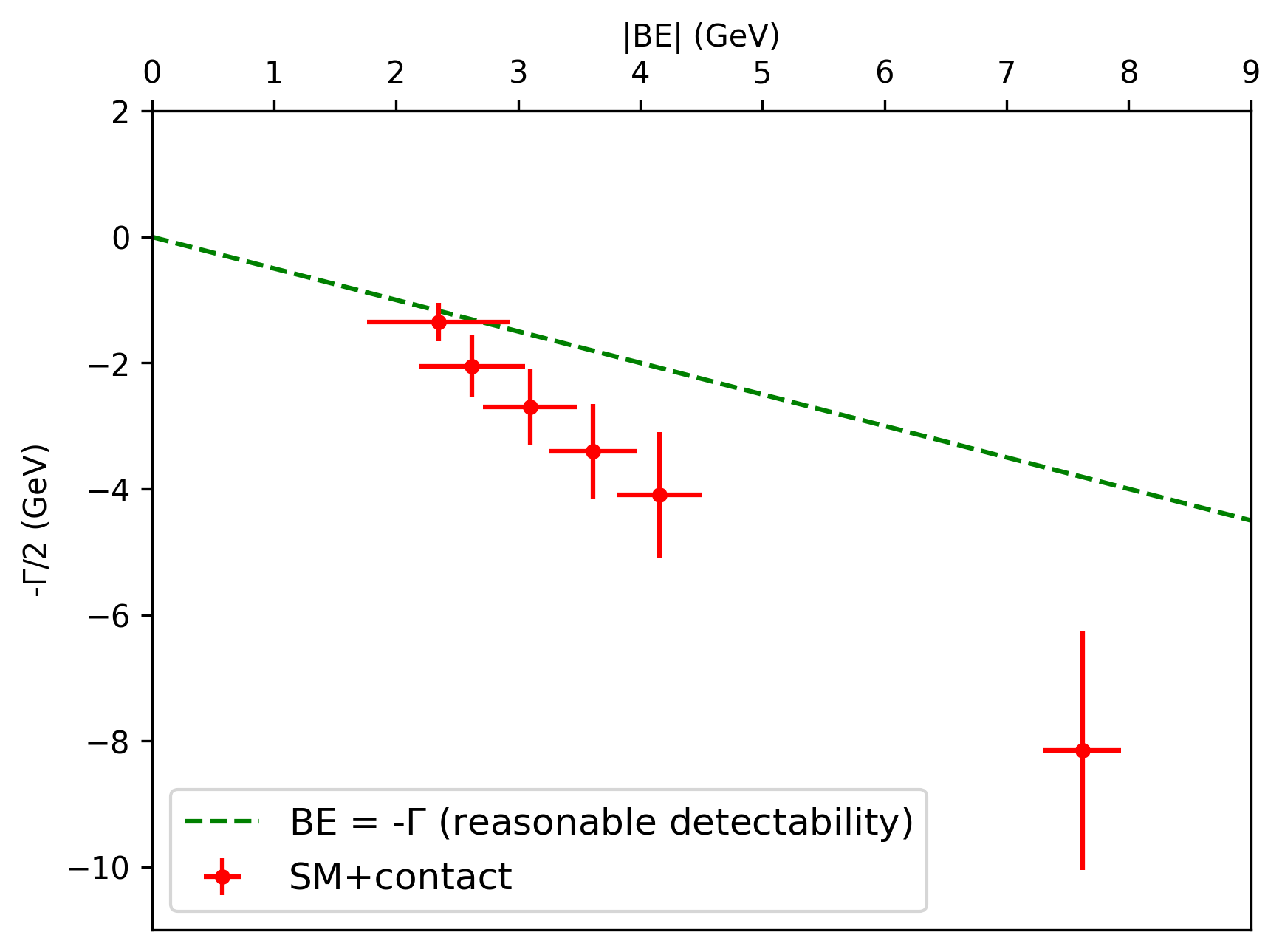}
    \caption{Multi-$t$ resonances (OX axis: binding energy; OY axis: half-width, as would appear on the second Riemann sheet of a scattering amplitude). Computer data, from left to right for $N =$ \{2, 3, 4, 5, 6, 12\}. Green dotted line: illustration of Rayleigh's detectability criterion, $|BE|\simeq \Gamma$. 
    \label{fig:begamma}}
\end{figure}

In Figure \ref{fig:begamma}, the multi-top resonances produced by the SM for static t-quarks at NLO in pNRQCD and BSM interactions parameterized within this work are pictured in the complex energy plane $(|BE|,-\Gamma/2)$.
The SM+contact binding energies are close to the SM ones, reflecting that they are a correction on the SM physics, so we only draw the full computation including the BSM part. (One can argue that this comes about because we have set $\Lambda\sim 1$ TeV, a high scale for $t$-physics, but that is what experimental data seems to be indicating.) It should be noted that, within experimental constraints of Wilson coefficients (Table \ref{tab:operators_sb}), toponium could still be detectable as a broad resonance akin to the $f_0(500)$ in low-energy hadron physics, its binding energy being near its width. For larger $N$, the resonances would be so deep in the complex plane that Rayleigh's criterion would not be satisfied, thus, the resonances would likely not be detectable at all.

\section{Conclusion}
We have deployed the bound-state Hartree-Fock method for hadrons composed exclusively of heavy  ($c$, $b$) quarks. The method is not as precise as, for example, the diffusion Monte Carlo computation~\cite{Gordillo:2024blx} recently proposed for pentaquark states, but it is practical for near-term applications and easy to deploy across many different numbers of particles and flavours, in addition to being transparent to most physicists.

Filling the $1s$ orbital with up to six quarks and six antiquarks because of the three-colour and two-spin degrees of freedom allows to reach up to 12 particles. Given rule of thumb estimates for production cross-sections, it is not worth addressing (currently, at least) such large states for hadron spectroscopy, so we have computed masses up to and including the all-heavy dibaryon with six quarks.

We fix the $c$, $b$ quark masses with the well-measured conventional charmonia and bottomonia, then proceed to predict masses of all-heavy tetraquarks, baryons, pentaquarks and the dibaryon. Our results are compatible with the $(ccb-bbc)$ dibaryon being bound, in total analogy with the deuteron in nuclear physics. But note that there is no pion exchange in our calculations, based on the colour interactions. 
We have not considered configuration mixing between different $N$ sectors (such as tetraquark and baryon-antibaryon mixing~~\cite{Karliner:2024cql} coupling $N=4$ to $N=6$), which should be considered for more realistic applications in meson spectroscopy.

We have however visited that dodecaquark in our extension of the work to the $t$-quark. The same principles apply to it, except that it being at the energy frontier, one could question whether new-physics interactions would be at play.
Therefore we have organized the computation so as to be able to
constrain new-physics parameters, such as SMEFT/HEFT Wilson coefficients. 

While the multiquark spectroscopy with charm and bottom quarks will certainly come to fruition to some extent, already in existing experiments, but also in those of the upcoming High-Luminosity LHC, the multi-top resonances are too broad to be reasonably detectable in the Standard Model. In particular, we concur with earlier work that the dodecatop or T-ball does not bind with Higgs-exchange alone, and binds but is too broad to be detected when adding the colour interactions.

Nevertheless, it can be used to constrain BSM parameters.
This method requires experimental collaborations to show various $N$-top spectra (from the top decay products), and exploits the eventual absence of any resonant structure. Four BSM interactions have been considered, thus, only four values of $N$ are necessary to constraint their Wilson coefficients.

On the other hand, we can turn the computation around and use the extant bounds on those coefficients of SMEFT/HEFT to see how far toponium is from being a detectable resonance, and it is in the limit of detectability, with comparable binding energy  and  width. 

In addition, we have concluded that, due to large widths compared with small binding energies, multi-top resonances with $N>2$ will likely not be detected.

\acknowledgments
Supported by grants PRX23/00225 (estancias en el extranjero) and PID2022-137003NB-I00 and PID2022-136510NB-C31 of the Spanish MCIN/AEI
/10.13039/501100011033/; EU’s 824093 (STRONG2020); FPU21/04180 of the Spanish Ministry of Universities; and Universidad Complutense de Madrid under research group 910309 and the IPARCOS institute.


\end{document}